\documentclass[aps,pra,showpacs,superscriptaddress,twocolumn]{revtex4}

\usepackage{amsmath,graphicx}
\usepackage{rotating}

\makeatletter\g@addto@macro\UrlSpecials{\do\!{\newline}}\makeatother

\newcommand{\tr}[2][]{\textnormal{tr}#1{{\left\{#2\right\}}}}

\newcommand{\Eq}[2][Eq.~]{#1(\ref{eq:#2})}
\newcommand{\Fig}[2][Fig.~]{#1\ref{fig:#2}}
\newcommand{\one}{\boldsymbol{1}}
\newcommand{\RFandP}[4]{\small$\begin{array}{rl}%
   \mathit{#1}&{#2}\\{#3}&{#4}\end{array}$}
\newcommand{\Step}[1]{\par\medskip\noindent\textbf{Step~#1}\quad\ignorespaces}
\newcommand{\gap}{\rule{1pt}{0pt}}
\newcommand{\half}{\frac{1}{2}}
\newcommand{\thalf}{\tfrac{1}{2}}
\newcommand{\D}{\mathrm{d}}
\newcommand{\plus}{\mathnormal{+}}
\newcommand{\ten}[1]{$\times10^{-#1}$}
\newcommand{\Exp}[1]{\mathrm{e}^{\mbox{\footnotesize$#1$}}}
\newcommand{\column}[2][c]{{\left(\begin{array}{#1}#2\end{array}\right)}}
\newcommand{\repr}{\mathrel{\widehat{=}}}

\newcommand{\KeyWords}[1]{\newline\rule{0em}{16pt}%
  {\footnotesize{Keywords:\hfill\begin{minipage}[t]{358pt}#1\end{minipage}}}}

\DeclareMathAlphabet{\vecfont}{OT1}{cmr}{bx}{it}
\renewcommand{\vec}[1]{\vecfont{#1}}
 
\newcommand{\vsigma}{\boldsymbol{\sigma}} 

\begin{document}

\title{Very strong evidence in favor of quantum mechanics\\%
       and against local hidden variables from a Bayesian analysis}

\author{Yanwu \surname{Gu}}\email{a0129460@u.nus.edu}
\affiliation{Department of Physics, National University of Singapore, 2
  Science Drive 3, Singapore 117542, Singapore} 
\affiliation{Centre for Quantum Technologies, National University of
  Singapore, 3 Science Drive 2, Singapore 117543, Singapore} 

\author{Weijun \surname{Li}}\email{cqtlw@nus.edu.sg}
\affiliation{Centre for Quantum Technologies, National University of
  Singapore, 3 Science Drive 2, Singapore 117543, Singapore} 

\author{Michael \surname{Evans}}\email{mevans@utstat.utoronto.ca}
\affiliation{Department of Statistical Sciences, University of Toronto, %
  Toronto, Ontario M5S~3G3, Canada}

\author{Berthold-Georg Englert}\email{cqtebg@nus.edu.sg}
\affiliation{Centre for Quantum Technologies, National University of
  Singapore, 3 Science Drive 2, Singapore 117543, Singapore} 
\affiliation{Department of Physics, National University of Singapore, 2
  Science Drive 3, Singapore 117542, Singapore} 
\affiliation{MajuLab, CNRS-UCA-SU-NUS-NTU International Joint Unit, Singapore}

\date[]{Posted on the arXiv on 21 August 2018; updated on 22 January 2019}

\begin{abstract}
The data of four recent experiments --- conducted in Delft, Vienna, 
Boulder, and Munich with the aim of refuting nonquantum 
hidden-variables alternatives to the quantum-mechanical description 
--- are evaluated from a Bayesian perspective of what constitutes 
evidence in statistical data. 
We find that each of the experiments provides strong, or very strong, evidence
in favor of quantum mechanics and against the nonquantum alternatives. 
This Bayesian analysis supplements the previous non-Bayesian ones, which
refuted the alternatives on the basis of small p-values, but could not support 
quantum mechanics.
\end{abstract}


\maketitle

\section{Introduction}\label{sec:intro}
Four recent experiments in Delft~\cite{Delft}, Vienna~\cite{Vienna},
Boulder~\cite{Boulder}, and Munich~\cite{Munich} tested the variants of Bell's 
inequality~\cite{Bell:64} introduced by Clauser \textit{et al.}
\cite{Clauser+3:69} and Eberhard~\cite{Eberhard:93}.
The shared aim of these experiments was the refutation of descriptions in
terms of local hidden variables (LHV) that Bell and others had proposed as an
alternative to the description offered by quantum mechanics (QM).
Upon extracting small p-values from the respective data, with values between
$3.74\times10^{-31}$ (Vienna) and $0.039$ (Delft), each of the four groups of
scientists concluded that their data refute the LHV hypothesis.
Putting aside all other caveats about, objections against, and other issues
with the use of p-values~\cite{Evans:p-value,ASA:16, Benjamin+71:17}, 
let us merely note that the use of p-values can only make a case
against LHV but not in support of QM. 
Yet, a clear-cut demonstration that the data give evidence in favor of QM is
surely desirable. 

We present here an evaluation of the data of the four experiments that shows
that there is very strong evidence in favor of QM and also against LHV.
Our analysis does not rely on p-values or any other concepts of frequentist
statistics. 
We use Bayesian logic and measure evidence --- in favor of alternatives or
against them --- by comparing the posterior with the prior probabilities of 
the alternatives to be distinguished.

The basic notion is both simple and natural~\cite{Evans:15}:
If an alternative is more probable in view of the data than before acquiring
them, then the data provide evidence in favor of this alternative; and,
conversely, if an alternative is less probable after taking note of the data 
than before, then the data give evidence against this alternative. 

In our analysis, we only employ this principle of evidence and no
particular measure for quantifying the strength of the
evidence~\cite{strength}.
As it happens, all alternatives save one are extremely improbable in view of
the data so that the evidence in favor of the privileged alternative is
overwhelming, and a quantification of the strength of the evidence is not
needed here.

\begin{figure}
  \centering
  \includegraphics[viewport=55 670 230 790]{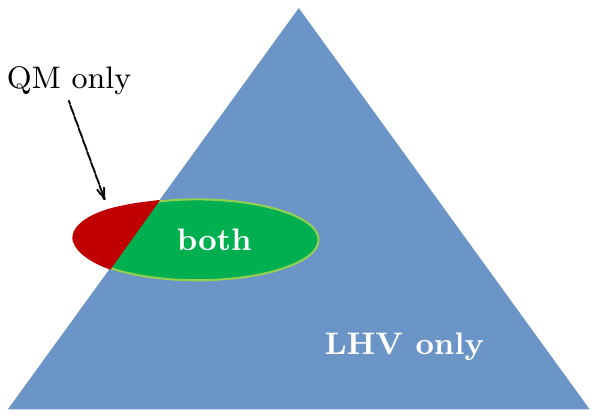}
  \caption{\label{fig:regions}%
      Symbolic sketch of the eight-dimensional set of permissible
      probabilities. 
      The points inside the ellipse symbolize probabilities accessible by
      quantum mechanics (QM); the triangle encloses the probabilities 
      permitted by local hidden variables (LHV).
      There are no QM probabilities in the blue portion of the LHV set, 
      and no LHV probabilities in the red part of the QM set.
      The green overlap region contains the probabilities that are possible
      both for QM and LHV.}
\end{figure}

While Bell's inequality and its variants are central to the design of the
experiments, they play no role in our evaluation of the data. 
What matters are the probabilities of occurrence of the various
measurement outcomes in the experiments.
As discussed in Sec.~\ref{sec:regions}, the permissible probabilities make up
an eight-dimensional set.
It is composed of three subsets: one accessible only by QM,
another only by LHV, and the third by both; see \Fig{regions}.
We then ask 
\textit{Do the data provide evidence in favor of or against each of the
three subsets?} 
and, from the data of each of the four experiments, 
we find strong evidence for the QM-only subset and against the other two.

An essential part of the Bayesian analysis is the choice of prior --- the
assignment of prior probabilities to the three regions in
\Fig{regions}, thereby accounting for our prior knowledge about the
experiment and the assumptions behind its design.
If we were to strictly follow the rules of Bayesian reasoning, we would endow
the set of QM-permitted probabilities with a prior close to 100\% and allocate
a very tiny prior to the subset of LHV-only probabilities.
For, generations of physicists have accumulated a very large body of solid
experimental and theoretical knowledge that makes us extremely confident that
QM is correct.
Not one observed effect contradicts the predictions of QM,
while there is not a single documented phenomenon in support of all those
speculations about LHV.
In fact, this was already the situation in the mid 1960s when Bell published
Ref.~\cite{Bell:64} and gave a physical interpretation to an inequality known
to Boole a century before~\cite{Boole:62}.

Moreover, for the evaluation of the data from the four experiments, a
by-the-rules prior, namely a properly elicited prior, would have to reflect
our strong conviction that the experimenters managed to implement the
experiment as planned in a highly reliable fashion, with the desired
probabilities from the QM-only subset. 
Accordingly, we really should assign a very large prior probability to the
``QM only'' region symbolized in \Fig{regions}, a much smaller one to the
``both'' region, and an even smaller one to the ``LHV only'' region.

Such a prior, however, could bias the data evaluation in
favor of QM and against LHV.
Therefore, we deliberately violate the rules and use a prior that treats QM
and LHV on equal footing; see Sec.~\ref{sec:prior}.
To demonstrate that our choice of prior is not biased toward QM, we check for
such a bias and confirm that there is none; more about this in
Sec.~\ref{sec:bias-check}.
Yet, all this tilting of the procedure does not help the LHV case: 
The data speak clearly and loudly that \emph{QM rules and LHV are out}.

This contributes also to the development of Bayesian methodology,
inasmuch as we demonstrate that the subjective biases inherent in a Bayesian
statistical analysis through, for example, the choice of the prior, can be
assessed a priori. 
To the best of our knowledge this is one of the first applications of this
type of computation to ensure that such choices are not producing foregone
conclusions.

We recall the experimental scheme common to all four experiments
(Sec.~\ref{sec:exp-scheme}) and  the ways in which the probabilities of
detecting the various events are parameterized in the QM formalism or by LHV
(Sec.~\ref{sec:regions}). 
This is followed by a discussion of how the difference between the prior and
the posterior content of a region gives evidence in favor of this region or
against it (Sec.~\ref{sec:evidence}).
Then we explain our choice of prior on the eight-dimensional set of
permissible probabilities --- permitted either by QM or by LHV
(Sec.~\ref{sec:prior}); more specifically, we define the prior by the
algorithm that yields the large sample of permissible probabilities needed for
the Monte Carlo integrations over the three regions symbolized in
\Fig{regions}.  

Then, having thus set the stage, we present, as a full illustration of the
reasoning and methodology,  the detailed account of the various aspects of our
evaluation of the data recorded in one run of the Boulder experiment
(Sec.~\ref{sec:Boulder-5}). 
This includes the estimation, from the data, of an experimental
parameter for which the value given in Ref.~\cite{Boulder} is not accurate.
While an accurate value is not needed for calculating the p-value
reported in Ref.~\cite{Boulder}, it is crucial for the QM account of the
experiment. 
The results of processing the data from three other runs of the Boulder
experiment are reported in Sec.~\ref{sec:Boulder-1or3or7}.
The evaluation of the data from three runs of the Vienna experiment also
requires the estimation of the analogous parameter (Sec.~\ref{sec:Vienna})
whereas there is no need for that in the context of the experiments conducted
in Delft and Munich (Sec.~\ref{sec:Delft+Munich}).

All four experiments separately provide strong evidence in favor of QM and
against LHV.
Jointly, they convey the very clear message that this verdict is final.

\section{Experimental scheme}\label{sec:exp-scheme}
\begin{figure}
  \centering
  \includegraphics{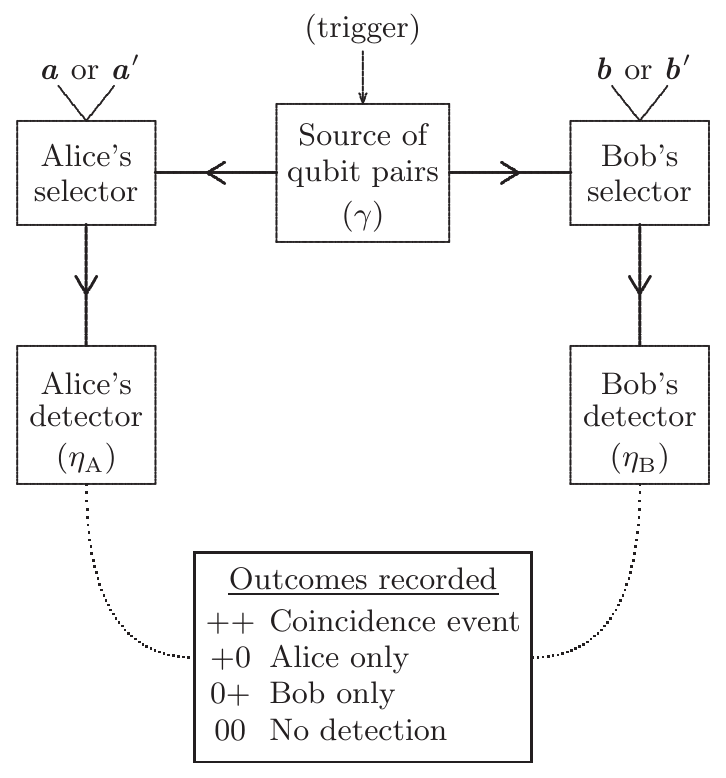}
  \caption{\label{fig:exp}Upon receiving a trigger signal,
   the source of qubit pairs equips Alice and Bob with one qubit each;
   the success probability for this initial step is $\gamma$.  
   The qubits are detected with the respective efficiencies 
   $\eta^{\ }_{\mathrm{A}}$ and $\eta^{\ }_{\mathrm{B}}$ if they pass a
   selection process, specified by $\vec{a}$ or $\vec{a}'$ for Alice's qubit
   and by $\vec{b}$ or $\vec{b}'$ for Bob's qubit.
   For each trigger signal, we record whether there was a coincidence event, or
   only Alice or only Bob observed a detector click, or both did not.
   Data are collected for all four settings:
   $\vec{a}$ or $\vec{a}'$ together with $\vec{b}$ or
   $\vec{b}'$.}
\end{figure}

The four experiments realize variations of one theme; see \Fig{exp}.
Upon receiving a trigger signal, the source of qubit pairs equips Alice and
Bob with one qubit each; 
the success probability for this is denoted by $\gamma$.
Alice chooses one of two settings, denoted by $\vec{a}$ and $\vec{a}'$, for
her selector in front of her qubit detector, which fires with efficiency
$\eta^{\ }_{\mathrm{A}}$.
Likewise Bob chooses between settings $\vec{b}$ and $\vec{b}'$ for his
selector and then detects the selected qubits with efficiency 
$\eta^{\ }_{\mathrm{B}}$.
For each trigger signal, the outcome is recorded and counts as an event of one
of four kinds: a ``$++$ event'' if Alice's and Bob's detectors both fire; 
a ``$+0$ event'' if Alice's detector fires and Bob's does not; 
a ``$0+$ event'' if Alice's detector does not fire and Bob's does; 
or a ``$00$ event'' if both detectors do not fire.
The data $D$ consist of the number of events observed of the four kinds, for the
four settings available by choosing $\vec{a}$ or $\vec{a}'$ and $\vec{b}$ or
$\vec{b}'$; together there are sixteen counts, such as $n^{(ab')}_{0+}$ for
the $0+$ events in the setting with $\vec{a}$ and $\vec{b}'$, and 
\begin{equation}
  \label{eq:A1}
  D=\Bigl(n^{(ab)}_{++},n^{(ab)}_{+0},\dots,
          n^{(ab')}_{0+},\dots,n^{(a'b')}_{00}\Bigr)
\end{equation}
reports the data for one run of the experiment as a 16-element string of
natural numbers.
Their sum
\begin{equation}
  \label{eq:A2}
  N=n^{(ab)}_{++}+n^{(ab)}_{+0}+\cdots+n^{(ab')}_{0+}+\cdots+n^{(a'b')}_{00}
\end{equation}
is the total number of trigger signals.

Table~\ref{tbl:4exp} lists the parameters of the four experiments
\cite{fn:data}. 
The Vienna and Boulder experiments exploit the polarization qubits of photon
pairs generated by down-conversion processes that happen rarely
($\gamma\ll1$).
The unit vectors $\vec{a}$, $\vec{a}'$ and $\vec{b}$, $\vec{b}'$ that specify
the selections refer to the orientation of polarization filters, and the
detectors register the photons that are let through.

The qubits in the Delft and Munich experiments are in superpositions of
hyperfine states of two atoms in spatially separated traps.
The preparation of the initial state is achieved by entanglement swapping and
is, therefore, heralded so that $\gamma=1$ in these event-ready setups.
The selection and detection are implemented by probing for a chosen
superposition, specified by the Bloch vectors $\vec{a}$, $\vec{a}'$ and
$\vec{b}$, $\vec{b}'$.

\begin{table}
  \centering
  \caption{\label{tbl:4exp}The parameters of the four experiments:
  the trigger--to--qubit-pair conversion probability $\gamma$;
  the angle $\theta^{\ }_{\mathrm{A}}$ between Alice's settings;
  the angle $\theta^{\ }_{\mathrm{B}}$ between Bob's settings;
  Alice's detection efficiency $\eta^{\ }_{\mathrm{A}}$;
  Bob's detection efficiency $\eta^{\ }_{\mathrm{B}}$;
  number $N$ of trigger signals for two runs (Delft, Munich), three runs
  (Vienna), or four runs (Boulder).}
  \begin{tabular}{@{}lcccccc@{}}
    \hline\hline
  Experiment & $\gamma$ & $\theta^{\ }_{\mathrm{A}}$ & $\theta^{\ }_{\mathrm{B}}$
                        & $\eta^{\ }_{\mathrm{A}}$ & $\eta^{\ }_{\mathrm{B}}$ 
                        & $N$\\
  \hline
    Delft \cite{Delft,Delft-2nd}
           & 1 & $90^\circ$ & $80.6^\circ$& 0.971& 0.963 & 245\\
           && && && 228\\[0.5ex]
  Vienna \cite{Vienna} & 0.003\gap5 & $64^\circ$ & $64^\circ$
                       & 0.786 & 0.762 & 3\gap843\gap698\gap536\\
           && && && 3\gap502\gap784\gap150\\
           && && && 9\gap994\gap696\gap192\\[0.5ex]
  Boulder \cite{Boulder} & 0.000\gap5 & $60.2^\circ$ & $60.2^\circ$
                        & 0.747      & 0.756 & 175\gap647\gap100\\
           && && && 527\gap164\gap272\\
           && && && 886\gap791\gap755\\
           && && && 1\gap244\gap205\gap032\\[0.5ex]
  Munich \cite{Munich,fn:Munich-eta} 
                         & 1 & $90^\circ$ & $90^\circ$ & 0.975 & 0.975
              &27\gap885\\ && && && 27\gap683\\
  \hline\hline
  \end{tabular}
\end{table}

Since the physics is independent of the coordinate systems adopted for the
description, we can regard $\vec{a}$ and $\vec{a}'$ as vectors in the $xz$
plane of the Bloch ball for Alice's qubit, and likewise $\vec{b}$ and
$\vec{b}'$ are in the $xz$ plane  for Bob's qubit.
All that matters is the angle $\theta^{\ }_{\mathrm{A}}$ between $\vec{a}$ and
$\vec{a}'$, and the angle $\theta^{\ }_{\mathrm{B}}$ between $\vec{b}$ and
$\vec{b}'$.
Our choice of coordinate systems is then such that
\begin{eqnarray}
  \label{eq:A3}
  {\left.\begin{array}{c}\vec{a} \\ \vec{a}'\end{array}\right\}}
  &=& \pm\vec{e}_x\sin\bigl(\thalf\theta^{\ }_{\mathrm{A}}\bigr)
      +\vec{e}_z\cos\bigl(\thalf\theta^{\ }_{\mathrm{A}}\bigr)\,,\nonumber\\
  {\left.\begin{array}{c}\vec{b} \\ \vec{b}'\end{array}\right\}}
  &=& \pm\vec{n}_x\sin\bigl(\thalf\theta^{\ }_{\mathrm{B}}\bigr)
      +\vec{n}_z\cos\bigl(\thalf\theta^{\ }_{\mathrm{B}}\bigr)\,,
\end{eqnarray}
where $\vec{e}_x$, $\vec{e}_y$, $\vec{e}_z$ are the cartesian unit vectors for
Alice's qubit and $\vec{n}_x$, $\vec{n}_y$, $\vec{n}_z$ are those for Bob's. 
Table~\ref{tbl:4exp} reports the respective values of 
$\theta^{\ }_{\mathrm{A}}$ and $\theta^{\ }_{\mathrm{B}}$.

A comment is in order about the table entries for $\theta^{\ }_{\mathrm{A}}$
and $\theta^{\ }_{\mathrm{B}}$. 
The information given in Refs.~\cite{Delft,Vienna,Boulder,Munich} is in terms
of angles with respect to a reference direction, such as the setting of a wave
plate relative to the conventional direction of vertical polarization.
This tells us, for each setting, the magnitudes of the probability amplitudes
in the superposition of vertical and horizontal polarizations but not their
complex phases.
It appears that the experimenters assumed that there are no relative phases,
and this assumption leads to the values of $\theta^{\ }_{\mathrm{A}}$ and
$\theta^{\ }_{\mathrm{B}}$ in Table~\ref{tbl:4exp}.
When the assumption is not made, we get a range of values for some of the
$\theta^{\ }_{\mathrm{A}}$s and $\theta^{\ }_{\mathrm{B}}$s, such as
${25.9^\circ-4.2^\circ<\half\theta^{\ }_{\mathrm{A}}<25.9^\circ+4.2^\circ}$ in
the Boulder experiment.
It is possible to estimate $\theta^{\ }_{\mathrm{A}}$ and 
$\theta^{\ }_{\mathrm{B}}$ from the data (in a manner analogous to that of
Sec.~\ref{sec:gamma}) and we performed such an estimation for the Boulder data
in Table~\ref{tbl:BoulderData-5pulses} below, with the outcome that 
$\theta^{\ }_{\mathrm{A}}=60.2^\circ$ is reasonable.
We regard this as assurance that there are no relative phases to be concerned
about.  

Another comment concerns the uncertainties of the
$\eta^{\ }_{\mathrm{A}}$s and $\eta^{\ }_{\mathrm{B}}$s, and of the 
$\theta^{\ }_{\mathrm{A}}$s and $\theta^{\ }_{\mathrm{B}}$s in
Table~\ref{tbl:4exp}, such as $\eta^{\ }_{\mathrm{A}}=74.7\pm0.3\%$ 
and $\eta^{\ }_{\mathrm{B}}=75.6\pm0.3\%$ for the Boulder
experiment~\cite{Boulder}. 
While the evaluation of the data reported in Secs.~\ref{sec:Boulder} and
\ref{sec:3others} refers to the parameter values in Table~\ref{tbl:4exp}, we
also used slightly different values for comparison and found that our
conclusions are not affected at all.

\section{Permissible probabilities}\label{sec:regions}
For each setting $S=ab$ or $ab'$ or $a'b$
or $a'b'$, we have the four probabilities
$p^{(S)}_{++}$, $p^{(S)}_{+0}$, $p^{(S)}_{0+}$, and $p^{(S)}_{00}$ of
recording the respective events for the next trigger signal.
These have unit sum,
\begin{equation}
  \label{eq:B1}
  \sum_{\alpha,\beta=+,0}p^{(S)}_{\alpha\beta}=1\,,
\end{equation}
but are otherwise unrestricted.
Accordingly, the quartets of probabilities for one setting $S$ compose the
whole standard probability 3-simplex.
There are four simplices of this kind, one for each setting, which are
linked because Alice's detection probabilities do not depend on Bob's
setting,
\begin{eqnarray}\label{eq:B2}
  p_+^{(a)}&\equiv&p^{(ab)}_{++}+p^{(ab)}_{+0}=p^{(ab')}_{++}+p^{(ab')}_{+0}\,,
\nonumber\\
  p_+^{(a')}&\equiv&p^{(a'b)}_{++}+p^{(a'b)}_{+0}=p^{(a'b')}_{++}+p^{(a'b')}_{+0}\,,
\end{eqnarray}
and Bob's detection probabilities do not depend on Alice's setting,
\begin{eqnarray}\label{eq:B3}
  p_+^{(b)}&\equiv&p^{(ab)}_{++}+p^{(ab)}_{0+}=p^{(a'b)}_{++}+p^{(a'b)}_{0+}\,,
\nonumber\\
  p_+^{(b')}&\equiv&p^{(ab')}_{++}+p^{(ab')}_{0+}=p^{(a'b')}_{++}+p^{(a'b')}_{0+}\,.
\end{eqnarray}
These are the so-called ``no signaling'' conditions.

Together, then, the sixteen probabilities $p^{(S)}_{\alpha\beta}$ obey eight
constraints and, therefore, the probability space is eight-dimensional.
The regions sketched in \Fig{regions} are regions in this
eight-dimensional probability space.
It is fully parameterized by the four probabilities in 
\Eq[Eqs.~]{B2} and \Eq[]{B3} and the four null-event probabilities
$p^{(S)}_{00}$, as illustrated by
\begin{eqnarray}\label{eq:B4}
  p^{(ab)}_{++}&=&p_+^{(a)}+p_+^{(b)}+p^{(ab)}_{00}-1 \,,\nonumber\\
  p^{(ab)}_{+0}&=&1-p^{(ab)}_{00}-p_+^{(b)}  \,,\nonumber\\
  p^{(ab)}_{0+}&=&1-p^{(ab)}_{00}-p_+^{(a)}  
\end{eqnarray}
for setting $S=ab$.

\subsection{QM probabilities}\label{sec:QM-p}
In the description offered by QM, a trigger signal results in a qubit pair
with probability $\gamma$ and yields nothing with probability $1-\gamma$.
With $\rho$ denoting the statistical operator of the qubit pair, 
$\vsigma$ the Pauli vector operator of a qubit, and $\one$ the identity
operator, we have
\begin{eqnarray}
  \label{eq:B5}
  p^{(a)}_+&=&\gamma\eta^{\ }_{\mathrm{A}}
        \tr{\thalf(\one+\vec{a}\cdot\vsigma)\otimes\one\,\rho}\,,\nonumber\\
  p^{(a')}_+&=&\gamma\eta^{\ }_{\mathrm{A}}
       \tr{\thalf(\one+\vec{a}'\cdot\vsigma)\otimes\one\,\rho}\,,\nonumber\\
  p^{(b)}_+&=&\gamma\eta^{\ }_{\mathrm{B}}
       \tr{\one\otimes\thalf(\one+\vec{b}\cdot\vsigma)\,\rho}\,,\nonumber\\
  p^{(b')}_+&=&\gamma\eta^{\ }_{\mathrm{B}}
            \tr{\one\otimes\thalf(\one+\vec{b}'\cdot\vsigma)\,\rho}
\end{eqnarray}
for Alice's and Bob's individual probabilities~\cite{Munich-dark}, and
the null-event probabilities are 
\begin{eqnarray}
  \label{eq:B6}
  p^{(ab)}_{00}
  &=&\gamma\,\mathrm{tr}\Bigl\{
 \bigl[\one-\eta^{\ }_{\mathrm{A}}\thalf(\one+\vec{a}\cdot\vsigma)\bigr]
  \nonumber\\&&\hphantom{\gamma\mathrm{tr}\Bigl\{}\otimes
 \bigl[\one-\eta^{\ }_{\mathrm{B}}\thalf(\one+\vec{b}\cdot\vsigma)\bigr]
  \rho\Bigr\}+(1-\gamma)\quad
\end{eqnarray}
and analogous expressions for $p^{(ab')}_{00}$, $p^{(a'b)}_{00}$, and
$p^{(a'b')}_{00}$.

Owing to their small $\gamma$ values, the probabilities in the Vienna and
Boulder experiments occupy only a very small portion of the linked 3-simplices
because the probabilities are bounded by 
\begin{eqnarray}
  \label{eq:B7}
  &&p^{(S)}_{++}\leq\gamma\eta^{\ }_{\mathrm{A}}\eta^{\ }_{\mathrm{B}}\,,\quad
    p^{(S)}_{00}\geq1-\gamma\,,\nonumber\\
  &&p^{(a)}_+,p^{(a')}_+\leq\gamma\eta^{\ }_{\mathrm{A}}\,,\quad
    p^{(b)}_+,p^{(b')}_+\leq\gamma\eta^{\ }_{\mathrm{B}}\,.
\end{eqnarray}
For the Delft and Munich experiments, which have $\gamma=1$ and
$\eta^{\ }_{\mathrm{A}},\eta^{\ }_{\mathrm{B}}\lesssim1$, no major portions of
the 3-simplices are excluded.

The set of permissible QM probabilities, enclosed by the symbolic ellipse in
\Fig{regions}, is made up by the probabilities
$p^{(S)}_{\alpha\beta}$ obtained from all thinkable $\rho$s in accordance with
\Eq[Eqs.~]{B4}--\Eq[]{B6}. 
Each statistical operator $\rho$ is represented by a density matrix --- 
a hermitian, nonnegative, unit-trace $4\times4$ matrix.
As a consequence of \Eq{A3}, the $p^{(S)}_{\alpha\beta}$s are linear
combinations of the expectation values of the eight operators
\begin{eqnarray}
  \label{eq:B8}
  &&\sigma_x\otimes\one\,,\quad  \sigma_z\otimes\one\,,\quad
    \one\otimes\sigma_x\,,\quad  \one\otimes\sigma_z\,,\nonumber\\
  &&\sigma_x\otimes\sigma_x\,,\quad  \sigma_z\otimes\sigma_x\,,\quad
    \sigma_x\otimes\sigma_z\,,\quad  \sigma_z\otimes\sigma_z\,,\qquad
\end{eqnarray}
all represented by real $4\times4$ matrices if we employ the standard real
$2\times2$ matrices for $\sigma_x$ and $\sigma_z$.
Therefore, only the real parts of the matrix elements of $\rho$ matter, and we
only need to consider $\rho$s represented by real density matrices, which
make up a nine-dimensional convex set.
The ninth parameter is the expectation value of 
$\sigma_y\otimes\sigma_y=(\sigma_x\otimes\sigma_z)(\sigma_z\otimes\sigma_x)$.

\subsection{LHV probabilities}\label{sec:LHV-p}
In the LHV reasoning, the sequence ``first create a qubit pair,
then select, finally detect'' of Sec.~\ref{sec:QM-p} is meaningless;
all that has meaning is ``detection event after trigger signal''
\cite{Brunner+4:14}.
There are no sequential processes controlled by the hidden variables
step-by-step, they control the overall process. 
Therefore, the trigger-to-pair probability $\gamma$ and the detection
probabilities $\eta^{\ }_{\mathrm{A}}$, $\eta^{\ }_{\mathrm{B}}$, which are
central to the correct application of Born's rule in \Eq[Eqs.~]{B5} and
\Eq[]{B6}, play no role when relating the $p^{(S)}_{\alpha\beta}$s to hidden
variables.  

Following Wigner~\cite{Wigner:70} and others (see, for example,
Refs.~\cite{Fine:82,Kaszlikowski:00}), 
we parameterize the LHV probabilities in terms of sixteen hypothetical
probabilities $w(\alpha\alpha'\beta\beta')$ from the 15-simplex,
\begin{equation}\label{eq:B9}
  \mathop{\sum_{\alpha,\alpha'=+,0}}_{\beta,\beta'=+,0}
  \!\!w(\alpha\alpha'\beta\beta')=1\quad
\mbox{with}\enskip w(\alpha\alpha'\beta\beta')\geq0\,.
\end{equation}
Here, $w(\alpha\alpha'\beta\beta')$ is the fictitious joint probability of
obtaining, for the next trigger signal, 
result $\alpha$ for Alice's setting $\vec{a}$, and
result $\alpha'$ for her setting $\vec{a}'$, and
result $\beta$ for Bob's setting $\vec{b}$, and also
result $\beta'$ for his setting $\vec{b}'$ 
(never mind that she has setting $\vec{a}$ \emph{or} $\vec{a}'$
and he has $\vec{b}$ \emph{or} $\vec{b}'$).
The eight marginal probabilities
\begin{eqnarray}\label{eq:B10}
  p^{(a)}_+&=&\sum_{\alpha',\beta,\beta'}\!w(\plus\alpha'\beta\beta')\,,\quad
  p^{(a')}_+ =\sum_{\alpha,\beta,\beta'}\!w(\alpha\plus\beta\beta')\,,\nonumber\\
  p^{(b)}_+&=&\sum_{\alpha,\alpha',\beta'}\!w(\alpha\alpha'\plus\beta')\,,\quad
  p^{(b')}_+ =\sum_{\alpha,\alpha',\beta}\!w(\alpha\alpha'\beta\plus)\,,
      \nonumber\\  
  p^{(ab)}_{00}&=&\sum_{\alpha',\beta'}w(0\alpha'0\beta')\,,\quad
  p^{(ab')}_{00} =\sum_{\alpha',\beta}w(0\alpha'\beta0)\,,\nonumber\\
  p^{(a'b)}_{00}&=&\sum_{\alpha,\beta'}w(\alpha00\beta')\,,\quad
  p^{(a'b')}_{00} =\sum_{\alpha,\beta}w(\alpha0\beta0)
\end{eqnarray}
then determine the sixteen $p^{(S)}_{\alpha\beta}$s in accordance with \Eq{B4}.
 
The hidden probabilities $w(\alpha\alpha'\beta\beta')$ control all aspects of
the experiment, and they are such that they mislead us into regarding QM
as correct. 
That is, the LHV probabilities $p^{(S)}_{\alpha\beta}$ should have as many
properties of the QM probabilities as possible. 
Therefore, we require that all inequalities in \Eq{B7} are respected by the LHV
probabilities. 
Through these inequalities, then, the values of 
$\eta^{\ }_{\mathrm{A}}$, $\eta^{\ }_{\mathrm{B}}$, and $\gamma$, which are
properties of the experimental apparatus, enter the LHV formalism.
Accordingly, the set of permissible LHV probabilities, enclosed by the
symbolic triangle in \Fig{regions}, is made up by the probabilities
$p^{(S)}_{\alpha\beta}$ obtained from all thinkable $w(\alpha\alpha'\beta\beta')$s
in accordance with \Eq[Eqs.~]{B4} and \Eq[]{B10}, subject to the constraints
in \Eq{B7}. 

Note that, as a consequence of the different values of 
$\eta^{\ }_{\mathrm{A}}$, $\eta^{\ }_{\mathrm{B}}$, and $\gamma$, we have
different sets of permissible probabilities for the four experiments.
Symbolically, there are several different ellipses and several different
triangles in \Fig{regions}.

Note also that the ``LHV only'' region is not empty.
For example, if we choose $w(\alpha\alpha'\beta\beta')=0$ for all hidden
probabilities except for
$w(\plus\plus\plus\plus)=\gamma\eta^{\ }_{\mathrm{A}}\eta^{\ }_{\mathrm{B}}$
and $w(0000)=1-\gamma\eta^{\ }_{\mathrm{A}}\eta^{\ }_{\mathrm{B}}$,
then the constraints of \Eq{B7} are obeyed and we get
$p_{++}^{(S)}=\gamma\eta^{\ }_{\mathrm{A}}\eta^{\ }_{\mathrm{B}}$
for all four settings; there is no statistical operator $\rho$ for which
the QM probabilities of \Eq[Eqs.~]{B5} and \Eq[]{B6} are like this.
And any $\rho$ for which a Bell-type inequality, such as
$p^{(ab')}_{+0}+p^{(a'b)}_{0+}+p^{(a'b')}_{++}\geq p^{(ab)}_{++}$, is violated
will give probabilities in the ``QM only'' region.

\section{Prior and posterior content; evidence}\label{sec:evidence}
We write $(\D p)\,w_0(p)$ for the prior probability assigned to the
infinitesimal vicinity  of a point 
$p=\bigl(p^{(ab)}_{++},\dots,p^{(a'b')}_{00}\bigr)$ in the probability space,
where the differential element
\begin{equation}\label{eq:C1}
  (\D p)=\D p^{(ab)}_{++}\,\D p^{(ab)}_{+0}\,\cdots\D p^{(a'b')}_{00}\,
         w^{\ }_{\mathrm{cstr}}(p)
\end{equation}
incorporates the constraints that restrict $p$ to the set of permissible
values, symbolized by the union of the regions enclosed by the ellipse and the
triangle in \Fig{regions}.
In particular, we have $w^{\ }_{\mathrm{cstr}}(p)=0$ when
\Eq[Eqs.~]{B1}--\Eq[]{B3} are not obeyed. 
Other constraints result from the nonnegativity of the statistical operator and
the hidden probabilities, and from the restrictions imposed by \Eq{B7}.
Although there are algorithms for checking whether the constraints are 
obeyed by any given $p$, we do not have an explicit expression for 
$w^{\ }_{\mathrm{cstr}}(p)$; we also do not need one. 

While $w^{\ }_{\mathrm{cstr}}(p)$ depends on the parameters of the
experiments, with different constraints for the four experiments because they
differ in the values of $\theta^{\ }_{\mathrm{A}}$, $\theta^{\ }_{\mathrm{B}}$, 
$\eta^{\ }_{\mathrm{A}}$, $\eta^{\ }_{\mathrm{B}}$, and $\gamma$ (see
Table~\ref{tbl:4exp}), the factor $w_0(p)$ reflects what we know about the
experiments before the data are acquired.
Our choice for $w_0(p)$ is discussed in Sec.~\ref{sec:prior};
here, we shall assume that a certain choice has been made.

Then
\begin{equation}
  \label{eq:C2}
  S_{\mathcal{R}}=\int\limits_{\mathcal{R}}(\D p)\,w_0(p)
\end{equation}
is the prior content of region $\mathcal{R}$ (its ``size'').
The three regions of interest are the ones symbolized by the red, blue, and
green areas in \Fig{regions}, that is: the sets of probabilities permitted
only by QM, only by LHV, or by both.
The three prior contents have unit sum,
\begin{equation}
  \label{eq:C3}
  S_{\mathrm{QM\,only}}^{\ }+S_{\mathrm{LHV\,only}}^{\ }+S_{\mathrm{both}}^{\ }=1\,,
\end{equation}
which states the normalization of $w_0(p)$ to unit integral.

The likelihood function $L(D|p)$ tells us how likely are the data 
$D=\bigl(n^{(ab)}_{++},\dots,n^{(a'b')}_{00}\bigr)$ if the probabilities 
$p$ are the case.
Since successive trigger signals and the resulting detection events are
statistically independent~\cite{runs-test}, the likelihood has the multinomial
form
\begin{eqnarray}\label{eq:C4}
  L(D|p)=\frac{N!}{4^N}\prod_{S,\alpha,\beta}
         \frac{{p^{(S)}_{\alpha\beta}}^{n^{(S)}_{\alpha\beta}}}
              {n^{(S)}_{\alpha\beta}!}\,, 
\end{eqnarray}
where we assume that the four settings are chosen randomly with equal
probability, and there is a new setting for each trigger signal.
The joint probability of having $p$ inside the region $\mathcal{R}$ and
observing the data $D$ is
\begin{equation}
  \label{eq:C5}
  \int\limits_{\mathcal{R}}(\D p)\,w_0(p)\,L(D|p)=L(D)\,C_{\mathcal{R}}(D)\,,
\end{equation}
where
\begin{equation}
  \label{eq:C6}
  L(D)=\int\limits_{\mathrm{all}}(\D p)\,w_0(p)\,L(D|p)
\end{equation}
is the overall probability of obtaining the data $D$, and $C_{\mathcal{R}}(D)$
is the conditional probability that $p$ is inside the region $\mathcal{R}$
given the data $D$.
There is \emph{evidence in favor of} the region $\mathcal{R}$ when 
$C_{\mathcal{R}}(D)>S_{\mathcal{R}}$, and there is \emph{evidence against} the
region $\mathcal{R}$ when  $C_{\mathcal{R}}(D)<S_{\mathcal{R}}$
\cite{Evans:15}.

In this posterior content $C_{\mathcal{R}}(D)$ of the region $\mathcal{R}$
(its ``credibility''),
\begin{equation}
  \label{eq:C7}
  C_{\mathcal{R}}=\frac{1}{L(D)}\int\limits_{\mathcal{R}}(\D p)\,w_0(p)\,L(D|p)
              =\int\limits_{\mathcal{R}}(\D p)\,w_D(p)\,,
\end{equation}
we recognize the posterior density
\begin{equation}
  \label{eq:C8}
  w_D(p)=\frac{w_0(p)\,L(D|p)}{L(D)}\,,
\end{equation}
the Bayesian update of the prior density $w_0(p)$ in the face of the data $D$.
The posterior contents of the three particular regions of interest also add up
to unity,
\begin{equation}
  \label{eq:C9}
  C_{\mathrm{QM\,only}}^{\ }+C_{\mathrm{LHV\,only}}^{\ }+C_{\mathrm{both}}^{\ }=1\,,
\end{equation}
as $w_D(p)$ is normalized, too.
Owing to the unit sums in \Eq[Eqs.~]{C3} and \Eq[]{C9}, whatever the data,
there will be evidence in favor of one of the regions, and evidence against
another, and we can have evidence in favor of the third region or evidence
against it.

Regarding the $p$-independent combinatorial factor in \Eq{C4} we note
the following.
This particular combination of factorials refers to the situation in which one
takes data until $N$, the number of trigger signals, reaches a pre-chosen
value. 
Such is the stopping rule of the Delft and Munich experiments.
Other stopping rules have other combinatorial factors.
For example, in the Boulder and Vienna experiments, the value of
${n^{(ab)}_{++}+n^{(ab')}_{+0}+n^{(a'b)}_{0+}+n^{(a'b')}_{++}}$ is pre-chosen
and sets the stopping rule.
Further, the factor $4^N$ in the combinatorial factor does not apply when the
settings are not equally likely.
Other modifications are required if several consecutive events are recorded
before the setting changes, as is the case in the Boulder experiment; see
Sec.~\ref{sec:Boulder}. 

In the context of our investigation here, however, it does not matter what the
stopping rule is.
The combinatorial factor associated with the rule cancels in \Eq{C8} and is of
no further consequence.
Therefore, we shall use the combinatorial factor of \Eq{C4} for all datasets
we evaluate, irrespective of the actual stopping rule.
The lack of dependence on the stopping rule is characteristic of Bayesian
inferences generally.

\section{Choice of prior}\label{sec:prior}
We need to choose the prior density $w_0(p)$ in order to give specific meaning
to the integrals in \Eq[Eqs.~]{C2}, \Eq[]{C6}, and \Eq[]{C7}.
These eight-dimensional integrals are computed by Monte Carlo integration, for
which we need a large sample of permissible $p$s such that the number of sample
points in a region $\mathcal{R}$ is proportional to its prior content
$S_{\mathcal{R}}$. 
It is, therefore, expedient to define $w_0(p)$ by the sampling algorithm, and
this is what we do.

For the reasons mentioned in the Introduction, we shall not choose the prior
following the rules of proper Bayesian reasoning.
Instead, we opt for a prior that, under the ideal circumstances of perfect
detectors, does not distinguish between QM and LHV for a single setting $S$.

Our samples are composed of $10^6$ sets of probabilities generated from 
randomly chosen quantum states plus another $10^6$ sets from random LHV.
We employ two sampling algorithms, one for QM and the other for LHV, so that
half of our sample points are from the symbolic ellipse of \Fig{regions},
and the other half from the triangle.
When marginalized over the other three settings, the sample points inside the
3-simplex of the fourth setting have equal density for both algorithms under
the ideal circumstances of $\eta^{\ }_{\mathrm{A}}=\eta^{\ }_{\mathrm{B}}=1$,
with the same marginalized prior for each of the four 3-simplices.

After completing the QM sampling and the LHV sampling, described in
Secs.~\ref{sec:prior-QM} and \ref{sec:prior-LHV}, and confirming that there is
no hidden bias in the sample (see Sec.~\ref{sec:bias-check}), 
we have a suitable random sample of $2\times10^6$ points in the
space of permissible probabilities --- permitted either by QM or by LHV, that
is --- and we also know how many sample points are in the three regions of
interest. 
Put differently, we know the three prior contents that are added in \Eq{C3}.
Owing to the random process of sampling, the sample has fluctuations which
give rise to sampling errors in $S_{\mathrm{QM\,only}}^{\ }$,
$S_{\mathrm{LHV\,only}}^{\ }$, and $S_{\mathrm{both}}^{\ }$, and also in other
quantities computed by Monte Carlo integration with this sample. 
For the applications reported in Secs.~\ref{sec:Boulder} and
\ref{sec:3others}, however, we find that a sample with $2\times10^6$ entries is
large enough to ensure that the sampling errors do not affect the conclusions;
more about this in Sec.~\ref{sec:SamplingError}.

\subsection{QM contribution to the sample}\label{sec:prior-QM}
In all four experiments, the source yields the qubit pairs in an entangled
state of high purity, a very good approximation of the pure target state
that motivates the experimental effort --- the two-qubit state that requires
the smallest threshold detector efficiency (Vienna and Boulder) or leads to 
the strongest violation of the Bell-type inequality (Delft and Munich).
With this in mind, we produce the QM sample by the following five-step
procedure. 

\Step{1}
Draw four independent real numbers $x_1$, $x_2$, $x_3$, $x_4$ from a normal
distribution with zero mean and unit variance i.e., the probability
element is
\begin{equation}
  \label{eq:D0}
  \D x\,\frac{1}{\sqrt{2\pi}}\Exp{-\half x^2}\,;
\end{equation}
then
\begin{equation}
  \label{eq:D1}
  \varrho=\frac{1}{x_1^2+x_2^2+x_3^2+x_4^2}
          \column{x_1 \\ x_2 \\ x_3 \\ x_4}
          \column[cccc]{x_1 & x_2 & x_3 & x_4}
\end{equation}
is a real pure-state density matrix.
Repeat three times, thus producing $\varrho_1$, $\varrho_2$, $\varrho_3$,
and $\varrho_4$. 

\Step{2} 
Use the convex sum
\begin{equation}
  \label{eq:D2}
  \rho\repr(1-3\epsilon)\varrho_1+\epsilon(\varrho_2+\varrho_3+\varrho_4)
\end{equation}
with $\epsilon=0.001$ to make up the density matrix of a high-purity
full-rank statistical operator $\rho$.

\Step{3} 
Calculate the probabilities $p^{(S)}_{\alpha\beta}$ of \Eq[Eqs.~]{B5}
and \Eq[]{B6} and enter $p=\bigl(p^{(ab)}_{++},\dots,p^{(a'b')}_{00}\bigr)$
into the sample. 

\Step{4}
By checking if $p$ is inside Fine's
polytope~\cite{Fine:82,Froissard:81,Brunner+4:14}, or by any other method,
determine whether $p$ belongs to the ``QM only'' or the ``both'' set of
probabilities. 

\Step{5}
Repeat Steps 1--4 until the sample has $10^6$ entries.

\medskip

\noindent Some comments are in order: 
(i) The probability element in \Eq{D0} is such that the pure-state density
matrices of \Eq{D1} are uniformly distributed over the 3-sphere; put
differently, the distribution is uniform for the Haar measure on O(4).
Then, for each pure-state $\varrho$ of Step~1 
and each setting $S$, the marginal distribution on the 3-simplex has the prior
element 
\begin{equation}
  \label{eq:D3}
  \D p^{(S)}_{++}\,\D p^{(S)}_{+0}\,\D p^{(S)}_{0+}\,\D p^{(S)}_{00}
  \frac{\delta\bigl(p^{(S)}_{++}+p^{(S)}_{+0}+p^{(S)}_{0+}+p^{(S)}_{00}-1\bigr)}
       {\pi^2\gamma\sqrt{p^{(S)}_{++}p^{(S)}_{+0}p^{(S)}_{0+}
                         \bigl[p^{(S)}_{00}-(1-\gamma)\bigr]}}
\end{equation}
if $\eta^{\ }_{\mathrm{A}}=\eta^{\ }_{\mathrm{B}}=1$, where all factors in the
argument of the square root must be positive.
(ii) The value chosen for $\epsilon$ in Step~2 is a compromise.
Values that are much bigger result in a prior content of the ``QM only''
region that is too small to be useful; values that are much smaller, by
contrast, yield a sample of quantum states with unreasonably high purity.
That said, other small values of $\epsilon$ could be chosen in Step~2, or one
could determine small $\epsilon$s at random by a suitable lottery.

\subsection{LHV contribution to the sample}\label{sec:prior-LHV}
The algorithm for sampling from the LHV-permissible probabilities consists of
the following five steps.

\Step{1}
Draw sixteen independent positive numbers $y_1$, $y_2$, \dots, $y_{16}$ from a
$\Gamma\bigl(\tfrac{1}{8},1\bigr)$ distribution, i.e., the probability
element is
\begin{equation}
  \label{eq:D4}
  \D y\,\frac{y^{-\frac{7}{8}}}{\bigl(-\frac{7}{8}\bigr)!}\Exp{-y}\,;
\end{equation}
then put
\begin{eqnarray}
  \label{eq:D5}
  w(\plus\plus\plus\plus)&=&\gamma \frac{y_1}{Y}\,,\nonumber\\
  w(\plus\plus\plus0)&=&\gamma \frac{y_2}{Y}\,,\nonumber\\[-2ex]
  &\vdots&\nonumber\\
  w(000\plus)&=&\gamma \frac{y_{15}}{Y}\,,\nonumber\\
  w(0000)&=&(1-\gamma)+\gamma \frac{y_{16}}{Y}\,,\nonumber\\
\mbox{with}\quad Y&=&y_1+y_2+\cdots+y_{16}\,.
\end{eqnarray}
Repeat three times, thus producing $w_1(\alpha\alpha'\beta\beta')$, 
\dots, $w_4(\alpha\alpha'\beta\beta')$.

\Step{2}
With the same value of $\epsilon$ as in \Eq{D2}, use the convex sum
\begin{eqnarray}
  \label{eq:D6}
  w(\alpha\alpha'\beta\beta')&=&(1-3\epsilon)  w_1(\alpha\alpha'\beta\beta')
 +\epsilon\bigl[w_2(\alpha\alpha'\beta\beta')
\nonumber\\&&\mbox{}
+w_3(\alpha\alpha'\beta\beta')+w_4(\alpha\alpha'\beta\beta')\bigr]
\end{eqnarray}
for calculating the probabilities $p^{(S)}_{\alpha\beta}$ of \Eq{B10}.

\Step{3}
Enter $p=\bigl(p^{(ab)}_{++},\dots,p^{(a'b')}_{00}\bigr)$ into the sample if
the inequalities of \Eq{B7} are obeyed, and proceed to Step~4;
otherwise discard this $p$ and return to Step~1.

\Step{4}
Use the procedure described in Sec.~4.3 of Ref.~\cite{Seah+4:15}, or any other
method, to determine whether this $p$ belongs to the ``LHV only'' or the
``both'' set of probabilities.

\Step{5}
Repeat Steps 1--4 until the sample has $10^6$ entries.

\medskip

\noindent Here, too, some comments are in order:
(i) The probability element in \Eq{D4}, with the particular power
$-\frac{7}{8}$, is such that we get, for each setting
$S$, the same single-setting marginal distribution on the 3-simplex as for
the QM sampling, that is: \Eq{D3} applies to the LHV sample as well.
(ii) We include the constraint ${p^{(S)}_{00}\geq1-\gamma}$ into the
parameterization of the $w(\alpha\alpha'\beta\beta')$s in Step~1 rather than
into the acceptance or rejection procedure of Step~2, for the technical reason
that this gives us a much higher acceptance rate when ${\gamma\ll1}$ as is the
case for the Vienna and Boulder experiments.
(iii) Having ensured that the respective Steps~1 of the QM and the LHV
sampling give the same single-setting marginal distribution, we choose the
same $\epsilon$ in the mixing in the respective Steps~2 to keep the
single-setting distributions on equal footing.

\subsection{Checking the prior for bias}\label{sec:bias-check}
It is important to confirm that there is no bias in the prior that would
make us unfairly prefer one conclusion over the others.
For example, if we were to conclude regularly that there is evidence in favor
of the ``QM only'' region for data that are typical for $p$s in the ``both''
region, that would indicate a procedural bias for the ``QM only'' region.

Accordingly, our test for a bias proceeds as follows (see Sec.~4.6 in
Ref.~\cite{Evans:15}). 
We draw a random $p$ from the prior for the experiment in question and
simulate data for this ``true $p$'' for as many trigger signals as in
the experimental data.
The simulated data give evidence in favor of some regions and against others.
This is repeated for many such mock-true probabilities $p$, one thousand or
more for each of the three regions.

In our tests, we almost never get evidence in favor of the ``QM only''
region for true $p$s from another region when evaluating the data
from the experiments conducted in Boulder and Vienna
(Tables~\ref{tbl:BoulderData-nobias} and \ref{tbl:BoulderData-nobias1or3or7} in
Sec.~\ref{sec:Boulder},
Table~\ref{tbl:ViennaData-nobias} in Sec.~\ref{sec:Vienna}).  
Less rare are cases with evidence in favor of the ``both'' region for a
mock-true $p$ in the ``QM only'' region, but that is of no concern.
Owing to the much smaller counts of events in the Delft and Munich
experiments, for them it happens more often that we find evidence for the ``QM
only'' region for true $p$s in the ``both'' region, and even for true $p$s in
the ``LHV only'' region (Table~\ref{tbl:DelftMunichData-nobias} in
Sec.~\ref{sec:Delft+Munich}). 
This is understandable since statistical fluctuations in the simulated data
have a much larger chance of producing somewhat untypical data when the data
are few; indeed, such evidence for a ``wrong'' region occurs more often when
simulating the Delft experiment than the Munich experiment, which has more
than one-hundred times as many counts.
In summary, the bias checks establish that there is no procedural bias
in favor of the ``QM only'' region.

\makeatletter\global\advance\@colroom10pt\relax\set@vsize\makeatother

\section{The Boulder experiment}\label{sec:Boulder}
In the Boulder experiment~\cite{Boulder}, every one of the settings was
active for about $200\,$ns before a random switch to another (or the same)
setting occurred.
Pulses of short-wavelength light, $12.6\,$ns apart, were impinging on the
nonlinear crystal that generated down-converted photon pairs with a longer
wavelength.
The fifteen pulses per setting constitute a \emph{trial}, and a selected subset
of corresponding pulses from all trials make up the trigger signals of a
\emph{run}. 
When selecting one pulse only (the 6th), one gets the run with one trigger
signal per trial; likewise selecting three pulses (the 5th, 6th, and 7th)
yields the run with three trigger signals per trial; there are also runs with
five or seven trigger signals per trial, obtained by selecting the 4th to 8th
pulses or the 3rd to 9th pulses, respectively.
In the runs with three, five, or seven trigger signals per trial, then, there
is the same setting for this many consecutive events before the setting is
changed at random. 
Further, since the raw-data trials, of fifteen pulses each, are the same for
all four runs, these runs are not referring to independently collected data.
Roughly one third of the events in the run with three trigger signals per trial
are also contained in the run with one trigger signal per trial, and
correspondingly for the other runs.

\subsection{Trials with five trigger signals}\label{sec:Boulder-5}
We give here a detailed evaluation of the run with five trigger
signals per trial.
In total, there are $N=886\gap791\gap755$ trigger signals in this run
\cite{Boulder-run}; see
Table~\ref{tbl:BoulderData-5pulses} for the observed data and
Table~\ref{tbl:4exp} for the parameters of the experiment.

\makeatletter\global\advance\@colroom10pt\relax\set@vsize\makeatother

\begin{table}
\caption{\label{tbl:BoulderData-5pulses}
Boulder data: Recorded counts in the run with five trigger signals
per trial, for ${N=886\gap791\gap755}$ trigger signals~\cite{Boulder-run}.}
\centerline{\begin{tabular}{c@{\quad}c@{\quad}c@{\quad}c@{\quad}c}
\hline\hline\rule{0pt}{12pt}%
$S$ & $n_{++}^{(S)}$ & $n_{+0}^{(S)}$ & $n_{0+}^{(S)}$ & $n_{00}^{(S)}$\\
\hline 
$ab$ & 6\gap378 & 3\gap289 & 3\gap147 & 221\gap732\gap456\\
$ab^{\prime}$ & 6\gap794 & 2\gap825 & 23\gap230 & 221\gap686\gap486\\
$a^{\prime}b$ & 6\gap486 & 21\gap358 & 2\gap818 & 221\gap635\gap498\\
$a^{\prime}b^{\prime}$ & 106 & 27\gap562 & 30\gap000 & 221\gap603\gap322\\
\hline\hline
\end{tabular}}
\end{table}

The left part of Table~\ref{tbl:BoulderData-SC} summarizes our findings.
While  almost all of the prior is shared, roughly equally, between the
``both'' and ``LHV only'' regions, the ``QM only'' region contains merely
$6\times10^{-4}$ of the prior.  
This is a consequence of the detector efficiencies of about $75\%$ --- above,
but not far above, the threshold found by Eberhard~\cite{Eberhard:93}.
The posterior, by contrast, is entirely confined to the ``both'' region, so
that these data are inconclusive:
very strong evidence in favor of ``both'' and against ``QM only'' and also
against ``LHV only''.

\begin{table}[b]
\caption{\label{tbl:BoulderData-SC}%
  Boulder data: Prior and posterior contents
  of the three regions for the experimental data of Table
  \ref{tbl:BoulderData-5pulses}. 
  The value $\gamma=0.000\gap5$ is given in Ref.~\cite{Boulder},
  whereas $\gamma=0.000\gap722$  is the value estimated from the data.
  A posterior content of ``$0$'' indicates a number not exceeding the smallest
  floating-point format of the software code ($\simeq10^{-320}$), 
  and the entries ``$1$'' are to be understood accordingly.
  The prior contents have sampling errors that are discussed in
  Sec.~\ref{sec:SamplingError}.}  
\centerline{\begin{tabular}{cclcclc}
\hline\hline\rule{0pt}{0pt}%
 && \multicolumn{2}{c}{$\gamma=0.000\gap5$} 
 && \multicolumn{2}{c}{$\gamma=0.000\gap722$}\\
region &\qquad\quad& prior & posterior &\qquad\quad& prior & posterior\\ \hline
 QM only && 0.000\gap6  & 0 && 0.000\gap6   & 1\\
  both   && 0.502\gap6  & 1 && 0.502\gap5   & 0\\
LHV only && 0.496\gap9  & 0 && 0.496\gap9   & 0\\ 
\hline\hline
\end{tabular}}
\end{table}

\begin{table*}
\caption{\label{tbl:BoulderData-MLE}
Boulder data for $\gamma=0.000\gap5$ (top) and $\gamma=0.000\gap722$ (bottom): 
Each entry in this two-fold $4\times7$ table is a $2\times2$ subtable
reporting, for the respective setting $S$, the observed relative frequencies
(top left, italics) as well as the probabilities for the target state (top
right), for the QM-MLE (bottom left), and for the LHV-MLE (bottom right).
Note that the relative frequencies are not permissible probabilities; they do
not respect the no-signaling conditions in \Eq[Eqs.~]{B2} and \Eq[]{B3}.} 
\begin{tabular}{c@{\ \ }cccccccc}%
\hline\hline\rule{0pt}{12pt}%
&$S$ & $10^6p^{(S)}_{++}$ & $10^6p^{(S)}_{+0}$ %
& $10^6p^{(S)}_{0+}$ & $10^6p^{(a)}_+$ & $10^6p^{(a')}_+$ %
& $10^6p^{(b)}_+$ & $10^6p^{(b')}_+$\\ \hline%
&$ab$ %
& \RFandP{   28.76}{   20.49}{   32.53}{   27.78}%
& \RFandP{   14.83}{    9.66}{   12.59}{   15.43}%
& \RFandP{   14.19}{   10.03}{   12.07}{   14.53}%
& \RFandP{   43.60}{   30.16}{   45.12}{   43.21} &%
& \RFandP{   42.95}{   30.52}{   44.60}{   42.31} &%
\\[2.5ex]%
&$ab'$ %
& \RFandP{   30.64}{   21.88}{   25.74}{   29.68}%
& \RFandP{   12.74}{    8.27}{   19.38}{   13.53}%
& \RFandP{  104.77}{   68.05}{   96.09}{  105.40}%
& \RFandP{   43.38}{   30.16}{   45.12}{   43.21} & &%
& \RFandP{  135.41}{   89.93}{  121.83}{  135.07}%
\\[2.5ex]%
&$a'b$ %
& \RFandP{   29.26}{   21.88}{   23.74}{   28.59}%
& \RFandP{   96.35}{   66.98}{   89.10}{   96.32}%
& \RFandP{   12.71}{    8.64}{   20.86}{   13.72} &%
& \RFandP{  125.61}{   88.86}{  112.84}{  124.91}%
& \RFandP{   41.97}{   30.52}{   44.60}{   42.31} &%
\\[2.5ex]%
\begin{turn}{90}\makebox[0pt][l]{%
   \underline{\rule{23pt}{0pt}$\gamma=0.000\gap5$\rule{23pt}{0pt}}}\end{turn}
&$a'b'$ %
& \RFandP{    0.48}{    0.45}{    0.68}{    0.53}%
& \RFandP{  124.34}{   88.41}{  112.16}{  124.38}%
& \RFandP{  135.34}{   89.48}{  121.16}{  134.55} &%
& \RFandP{  124.82}{   88.86}{  112.84}{  124.91} &%
& \RFandP{  135.82}{   89.93}{  121.83}{  135.07}%
\\%
&&\multicolumn{7}{l}{\hrulefill}\\ \rule{0pt}{20pt}%
&$ab$ %
& \RFandP{   28.76}{   29.59}{   28.49}{   27.78}%
& \RFandP{   14.83}{   13.96}{   14.87}{   15.43}%
& \RFandP{   14.19}{   14.48}{   13.87}{   14.53}%
& \RFandP{   43.60}{   43.55}{   43.37}{   43.21} &%
& \RFandP{   42.95}{   44.07}{   42.36}{   42.31} &%
\\[2.5ex]%
&$ab'$ %
& \RFandP{   30.64}{   31.60}{   30.62}{   29.68}%
& \RFandP{   12.74}{   11.95}{   12.74}{   13.53}%
& \RFandP{  104.77}{   98.26}{  104.77}{  105.40}%
& \RFandP{   43.38}{   43.55}{   43.37}{   43.21} & &%
& \RFandP{  135.41}{  129.86}{  135.39}{  135.07}%
\\[2.5ex]%
&$a'b$ %
& \RFandP{   29.26}{   31.60}{   29.36}{   28.59}%
& \RFandP{   96.35}{   96.72}{   95.99}{   96.32}%
& \RFandP{   12.71}{   12.47}{   13.01}{   13.72} &%
& \RFandP{  125.61}{  128.32}{  125.35}{  124.91}%
& \RFandP{   41.97}{   44.07}{   42.36}{   42.31} &%
\\[2.5ex]%
\begin{turn}{90}\makebox[0pt][l]{%
   \underline{\rule{18pt}{0pt}$\gamma=0.000\gap722$\rule{18pt}{0pt}}}%
\end{turn}
&$a'b'$ %
& \RFandP{    0.48}{    0.65}{    0.50}{    0.53}%
& \RFandP{  124.34}{  127.67}{  124.85}{  124.38}%
& \RFandP{  135.34}{  129.21}{  134.89}{  134.55} &%
& \RFandP{  124.82}{  128.32}{  125.35}{  124.91} &%
& \RFandP{  135.82}{  129.86}{  135.39}{  135.07}%
\\%
\hline\hline%
\end{tabular}%
\end{table*}

This verdict is completely at odds with that reached by the authors of
Ref.~\cite{Boulder} who confidently reject the hypothesis of LHV on the basis
of their data.  
A careful consideration of all aspects of the experiment convinced us that the
discrepancy originates in the inaccurate value of the 
trigger-signal--to--qubit-pair conversion probability $\gamma$, given as
``$\,\approx5\times10^{-4}\,$'' in Ref.~\cite{Boulder}. 
When we use our best guess for $\gamma$ --- estimated from the data as
described below --- namely $\gamma=7.22\times10^{-4}$, which is some 40\%
larger than the quoted value, we get the numbers in the right part of
Table~\ref{tbl:BoulderData-SC}. 
While there is little change in the prior contents of the three regions, the
posterior is now entirely contained in the ``QM only'' region, so that we have
very strong evidence in favor of this region and against the other two,
against LHV that is.
Accordingly, we confirm that the LHV hypothesis is rejected, indeed.

It is worth noting here that $\gamma$ plays a very different role in the QM
formalism than in the LHV formalism.
The QM probabilities in \Eq[Eqs.~]{B5} and \Eq[]{B6} involve $\gamma$ quite
explicitly, whereas it restricts the LHV probabilities through the bounds in
\Eq{B7}.  
Therefore, a change in the value of $\gamma$ has quite different consequences
for the points of view offered by QM and LHV.
This is clearly demonstrated by the numbers reported in
Table~\ref{tbl:BoulderData-SC} and also by those in
Tables~\ref{tbl:BoulderData-MLE} and \ref{tbl:BoulderData-BA} 
as well as \Fig{MaxLik1} in the next section.

\subsubsection{Estimating $\gamma$ from the data}\label{sec:gamma}
As an exercise in quantum state estimation (QSE; see, for example,
Refs.~\cite{LNP649,Shang+4:13,Teo:16}), we determine the QM probabilities
$p^{(S)}_{\alpha\beta}$ that maximize the likelihood $L(D|p)$ of \Eq{C4} and
so find the QM-based maximum-likelihood estimator (QM-MLE; see
Ref.~\cite{Shang+2:17} for a fast and reliable algorithm).
Another maximization of $L(D|p)$, now over the LHV-permissible probabilities,
identifies the LHV-based maximum-likelihood estimator (LHV-MLE).
In the top part of Table~\ref{tbl:BoulderData-MLE}, we compare the probabilities
$p^{(S)}_{\alpha\beta}$ of the two MLEs with those of the target state --- the
ideal two-qubit quantum state that the source should make available --- and
with the relative frequencies associated with the counts in
Table~\ref{tbl:BoulderData-5pulses}. 
The $2\times2$ subtables are composed of the four corresponding probabilities,
with substantial variation within most of the subtables.

What is particularly unsettling is the colossal ratio of the maximum
values of the likelihood: $3.04\times10^{-712}$ (QM) versus
$2.29\times10^{-58}$ (LHV).
The data are much much more likely for LHV than for QM --- by more than $650$
orders of magnitude.
What is often termed ``the largest discrepancy in physics''~\cite{Adler+2:95},
a modest $120$ orders of magnitude, pales in comparison.

Now, the methods of QSE can be used for determining parameters of the
experiment in addition to the $p^{(S)}_{\alpha\beta}$s of the MLE.
One then speaks of \emph{self-calibrating QSE};
see, e.g., Refs.~\cite{Mogilevtsev:10,Branczyk+5:12,Quesada+2:13}.
In particular, one can optimize both the statistical operator $\rho$ of 
\Eq[Eqs.~]{B5} and \Eq[]{B6} and also the value of $\gamma$ when maximizing
the likelihood.
As stated above, the best guess we thus obtain is $\gamma=7.22\times10^{-4}$,
for which the maximum value of the likelihood is $2.55\times10^{-47}$ (QM);
the choice for $\gamma$ (in this range) has no effect on the LHV value of
$2.29\times10^{-58}$. 
For this optimized $\gamma$ value, then, the data are much more likely for QM
than for LHV --- by eleven orders of magnitude.

In passing we note that such small values of the likelihood are not surprising
if there are so many counts, simply because there is a huge number of similar
data, with a slight redistribution of counts, that could have been observed
equally well.
An absolute upper bound is given by the maximum of $L(D|p)$ over all
$p^{(S)}_{\alpha\beta}$s that obey the no-signaling constraints but are
otherwise unrestricted. 
This establishes $L(D|p)\leq2.90\times10^{-47}$, less than 15\% in excess of
the maximum of $L(D|p)$ over the QM-permissible probabilities~\cite{Signaling}.

The bottom part of Table~\ref{tbl:BoulderData-MLE} reports the probabilities
$p^{(S)}_{\alpha\beta}$ for ${\gamma=0.000\gap722}$. 
There is less variation within the subtables and, in particular, the relative
frequencies resemble the probabilities of the target state much better, and
also those of the QM-MLE.

This observation can be quantified, for which purpose we use (a variant of)
the Bhattacharyya angle~\cite{Bhattacharyya:43} between two sets of 
$p^{(S)}_{\alpha\beta}$s, computed by the following algorithm.
First, for each set of $p^{(S)}_{\alpha\beta}$s we introduce a corresponding
set of $q^{(S)}_{\alpha\beta}$s in accordance with
\begin{eqnarray}\label{eq:E1}
  p^{(S)}_{\alpha\beta}&=&4\gamma q^{(S)}_{\alpha\beta}
 \quad\mbox{if\ }\alpha\beta\neq00\,,\nonumber\\
   p^{(S)}_{00}&=&(1-\gamma)+4\gamma q^{(S)}_{00}\,;
\end{eqnarray}
the $q^{(S)}_{\alpha\beta}$s are positive and have unit sum,
\begin{equation}
  \label{eq:E2}
  \sum_{S}\sum_{\alpha,\beta} q^{(S)}_{\alpha\beta}=1\,,
\end{equation}
as implied by \Eq[Eqs.~]{B7} and \Eq[]{B1}.
Second, for any two sets of $p^{(S)}_{\alpha\beta}$s we compute the 
Bhattacharyya fidelity $F_{\mathrm{B}}^{\ }(p,p')$,
\begin{equation}
  \label{eq:E3}
  F_{\mathrm{B}}^{\ }(p,p')=\sum_{S}\sum_{\alpha,\beta}
       \sqrt{q^{(S)}_{\alpha\beta}{q'}^{(S)}_{\alpha\beta}}\,,
\end{equation}
and then the Bhattacharyya angle
\begin{equation}
  \label{eq:E4}
   \phi_{\mathrm{B}}^{\ }(p,p')=\cos^{-1}\bigl(F_{\mathrm{B}}^{\ }(p,p')\bigr)\,,
\end{equation}
whereby $0\leq F_{\mathrm{B}}\leq1$ and
$0\leq\phi^{\ }_{\mathrm{B}}\leq\half\pi$.
The smaller the value of $\phi_{\mathrm{B}}^{\ }(p,p')$, the more similar
are the two sets of probabilities.

\begin{table}
\caption{\label{tbl:BoulderData-BA} 
  Boulder data: Bhattacharyya angles
  between the observed relative frequencies and the probabilities of the
  target state and the two maximum-likelihood (ML) estimators.}  
\centerline{\begin{tabular}{l@{\qquad}c@{\qquad}c@{\quad}c}
\hline\hline
               & target     & \multicolumn{2}{c}{ML estimators}\\[-0.5ex]
$\quad\gamma$  & state      &   QM   & LHV \\ \hline
$0.000\gap5$   & $0.1164$ & $0.0462$ & $0.0056$\\
$0.000\gap722$ & $0.0106$ & $0.0014$ & $0.0047$\\ \hline\hline   
  \end{tabular}}
\end{table}

Table~\ref{tbl:BoulderData-BA} shows the Bhattacharyya angles between the
relative frequencies and the probabilities of the target state and the two
MLEs, both for ${\gamma=0.000\gap5}$ and for ${\gamma=0.000\gap722}$.
Clearly, the relative frequencies resemble the target-state probabilities and
the QM-MLE probabilities much better for $\gamma=0.000\gap722$ than for
$\gamma=0.000\gap5$; 
for the LHV-MLE probabilities, the difference between the angles for the two
$\gamma$ values is minimal and it originates entirely in the implicit $\gamma$
dependence of the Bhattacharyya angle that we introduce in \Eq{E1}.

\begin{figure}
  \centering
  \includegraphics{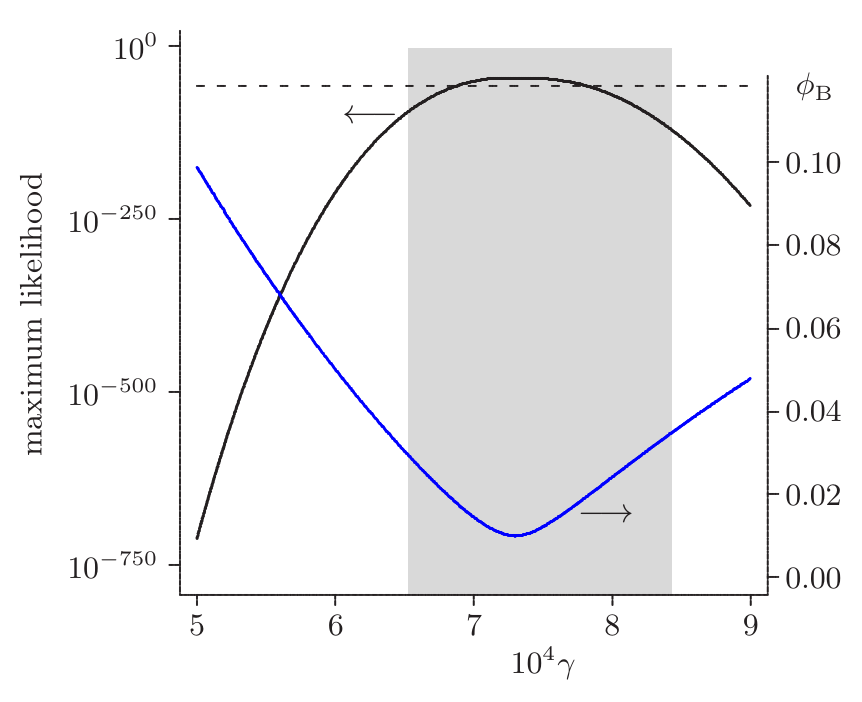}
  \caption{\label{fig:MaxLik1}%
    Boulder data: For ${0.000\gap5\leq\gamma\leq0.000\gap9}$, the solid curve
    in black graphs the maximum value of $L(D|p)$ over the QM-permitted
    probabilities (inside the symbolic ellipse in \Fig{regions}); the dashed
    black line marks the maximum value of $L(D|p)$ over the LHV-permitted
    probabilities (inside the symbolic triangle in \Fig{regions}); and the
    solid curve in blue is the graph of the Bhattacharyya angle 
    $\phi_{\mathrm{B}}^{\ }$ between the $p^{(S)}_{\alpha\beta}$s of the target
    state and those of the QM-MLE.}
\end{figure}

We close this discussion with a look at \Fig{MaxLik1}.
It shows, for $\gamma$ between $0.000\gap5$ and $0.000\gap9$, the maximum
value of the likelihood on the set of QM probabilities (solid black curve) and
on the set of LHV probabilities (dashed black line).
We observe that the QM value ranges over very many orders of magnitude while
the LHV value is independent of $\gamma$ in this interval.

This assures that there is no need for an accurate value of $\gamma$ if
one is only interested in the LHV description for the experiment when, for
example, refuting the LHV hypothesis on the basis of small p-values.
In our Bayesian evaluation of the data, however, we look for evidence in favor
of QM in addition to evidence against LHV, and the QM treatment of the data
requires an accurate value for $\gamma$.
It is fortunate that we can estimate $\gamma$ reliably from the data
themselves.

The grey strip in \Fig{MaxLik1} marks the $\gamma$ values for which the
probabilities of the QM-MLE violate a Bell inequality of the Eberhard
kind~\cite{Eberhard:93}; $\gamma=0.000\gap5$ is clearly outside.
This tells us once more that the actually observed data are typical for
$\gamma=0.000\gap722$ but not typical at all for $\gamma=0.000\gap5$
\cite{model-checking}.

The blue curve in \Fig{MaxLik1} shows the Bhattacharyya angle
$\phi_{\mathrm{B}}^{\ }$ between the $p^{(S)}_{\alpha\beta}$s of the target
state and those of the QM-MLE.
It acquires values between $0.010\gap0$ for $\gamma=0.000\gap73$ 
and $0.098\gap9$ for $\gamma=0.000\gap5$. 
By itself, this does not provide a reliable estimate for $\gamma$ but it is
consistent with, and so supports, our conclusion that $\gamma=0.000\gap722$ is
a value with much better justification than $\gamma=0.000\gap5$.

The graph of the maximum likelihood as a function of $\gamma$ in \Fig{MaxLik1}
shows a broad maximum.
This implies that values close to $\gamma=0.000\gap722$ can be chosen just as
well.  
Indeed, our conclusions are unchanged if slightly different $\gamma$ values
are used.
Should it be necessary, for another application, to make a quantitative
statement about the precision with which we infer the value of $\gamma$
from the data in Table~\ref{tbl:BoulderData-5pulses}, one could determine the
smallest credible intervals, among them the interval of plausible values,
with the methods described in Ref.~\cite{Li+3:16}.
This is, however, not worth the trouble in the present context.

\subsubsection{Sampling error}\label{sec:SamplingError}
In Sec.~\ref{sec:prior}, we mentioned that there are unavoidable sampling
errors. 
As the primary consequence, the prior contents of the three regions listed in
Table~\ref{tbl:BoulderData-SC} are uncertain.
Let us see to which extent. 

In the QM sampling of Sec.~\ref{sec:prior-QM}, the next $p$ has a probability
$a$ of belonging to the ``QM only'' region and a probability ${1-a}$ for
the ``both'' region.
The likelihood of getting a sample with $n_1$ $p$s in the ``QM only'' region
and $n_2$ $p$s in the ``both'' region is
\begin{eqnarray}\label{eq:E5}
L^{(\mathrm{QM})}_{\mathrm{spl}}(n_1,n_2|a)
&=&\frac{(n_1+n_2)!}{n_1!\;n_2!}a^{n_1}(1-a)^{n_2}\nonumber\\
&\leq&  L^{(\mathrm{QM})}_{\mathrm{spl}}(n_1,n_2|\hat{a})
\end{eqnarray}
which is largest for $\hat{a}=n_1/(n_1+n_2)$.
Analogously, the LHV sampling of Sec.~\ref{sec:prior-LHV} yields a sample with
$n_3$ $p$s in the ``LHV only'' region and $n_4$ $p$s in the ``both'' region
with a likelihood of
\begin{eqnarray}\label{eq:E6}
L^{(\mathrm{LHV})}_{\mathrm{spl}}(n_3,n_4|b)
&=&\frac{(n_3+n_4)!}{n_3!\;n_4!}b^{n_3}(1-b)^{n_4}\nonumber\\
&\leq&  L^{(\mathrm{LHV})}_{\mathrm{spl}}(n_3,n_4|\hat{b})
\end{eqnarray}
where $b$ is the probability that the next $p$ belongs to the ``LHV only''
region, and $\hat{b}=n_3/(n_3+n_4)$ is the maximum-likelihood estimator for
$b$.

Our best guesses, then, for the prior contents of the three regions are
\begin{eqnarray}\label{eq:E7}
&&S_{\mathrm{QM\,only}}^{\ }=\thalf\hat{a}\,,\quad
S_{\mathrm{LHV\,only}}^{\ }=\thalf\hat{b}\,,\nonumber\\
&&S_{\mathrm{both}}^{\ }=1-\thalf(\hat{a}+\hat{b})\,,  
\end{eqnarray}
since the two subsamples are of equal size, $n_1+n_2=n_3+n_4=10^6$.
Specifically, the sample for $\gamma=0.000\gap722$ has $n_1=1\gap210$ and
$n_3=993\gap805$, so that $\hat{a}=0.001\gap210$ and $\hat{b}=0.993\gap805$
with corresponding entries in Table~\ref{tbl:BoulderData-SC}.

With flat priors on $a$ and $b$, which are the quantities we need to infer
from the sampling frequencies, we have the posterior element
\begin{eqnarray}
  \label{eq:E8}
  \D a\,w^{(\mathrm{QM})}_{n_1,n_2}(a)
&=&\D a\,\frac{(n_1+n_2+1)!}{n_1!\;n_2!}a^{n_1}(1-a)^{n_2}\nonumber\\
&=&\D a\,w^{(\mathrm{QM})}_{n_1,n_2}(\hat{a})
   \Bigl(\frac{a}{\hat{a}}\Bigr)^{n_1}
   \Bigl(\frac{1-a}{1-\hat{a}}\Bigr)^{n_2}\nonumber\\
&\simeq&\D a\,w^{(\mathrm{QM})}_{n_1,n_2}(\hat{a})
        \,\Exp{-\half\frac{(a-\hat{a})^2}{v_a}}
\end{eqnarray}
for $a$ with the variance 
\begin{equation}
  \label{eq:E9}
v_a=\frac{\hat{a}(1-\hat{a})}{n_1+n_2}
   =\frac{n_1\,n_2}{(n_1+n_2)^3}\,,  
\end{equation}
and analogous expressions for $\D b\, w^{(\mathrm{LHV})}_{n_3,n_4}(b)$ and $v_b$.
Accordingly, the usual
one-standard-deviation confidence intervals for $a$ and $b$ are 
${\hat{a}-\sqrt{v_a}<a<\hat{a}+\sqrt{v_a}}$
and ${\hat{b}-\sqrt{v_b}<b<\hat{b}+\sqrt{v_b}}$, respectively.

More in the spirit of Bayesian inference than such confidence intervals, and
rather more conservative, is the plausible interval~\cite{plausible} which
consists of all values for which
\begin{equation}
  \label{eq:E10}
  L^{(\mathrm{QM})}_{\mathrm{spl}}(n_1,n_2|a)
  \geq\int\limits_0^1\hspace*{-0.2em}\D a'\,
  L^{(\mathrm{QM})}_{\mathrm{spl}}(n_1,n_2|a')
=\frac{1}{n_1+n_2+1}
\end{equation}
and analogously for $b$.
These intervals are 
\begin{eqnarray}\label{eq:E11}
  &&0.001\gap066<a<0.001\gap367\,,\nonumber\\
  &&0.993\gap475<b<0.994\gap124\,.
\end{eqnarray}
We thus arrive at
\begin{eqnarray}\label{eq:E12}
  &&0.000\gap533<0.000\gap588<S_{\mathrm{QM\,only}}^{\ }\nonumber\\
  &&\hphantom{0.000\gap533<0.000\gap593}%
    <0.000\gap622<0.000\gap683\,,\nonumber\\
  &&0.496\gap738<0.496\gap863<S_{\mathrm{LHV\,only}}^{\ }\nonumber\\
  &&\hphantom{0.496\gap738<0.496\gap501}%
    <0.496\gap942<0.497\gap062
\end{eqnarray}
for the one-standard-deviation intervals inside the plausible intervals.
These bounds quantify the sampling errors in the prior contents for
$\gamma=0.000\gap722$ in Table~\ref{tbl:BoulderData-SC}.
Very similar statements apply to the sample for $\gamma=0.000\gap5$ for which
$n_1=1\gap173$ and $n_2=993\gap713$.

The uncertainty of the numerical values of $S_{\mathrm{QM\,only}}$,
$S_{\mathrm{both}}$, and $S_{\mathrm{LHV\,only}}$ is sufficiently small to be
of no further concern.
Since \emph{all} of the posterior content is in the ``QM only'' region (for
$\gamma=0.000\gap722$), there is no doubt that the data give strong evidence
in favor of this region and against the other two.

\subsubsection{No bias in the prior}\label{sec:NoBias}
We check the sample of $p$s for a bias by the procedure described in
Sec.~\ref{sec:bias-check}, with the main objective of ensuring that the
sampling algorithm does not bias the outcome in favor of the ``QM only''
region. 
We generate ten thousand $p$s at random in each of the three regions and
simulate the Boulder experiment for these mock-true sets of probabilities. 
The resulting data are then examined whether they give evidence in favor of
one of the regions.
Table~\ref{tbl:BoulderData-nobias} summarizes this bias check, for
$\gamma=0.000\gap722$.

\begin{table}
  \caption{\label{tbl:BoulderData-nobias}%
     Boulder data, bias check of the prior:
     How often we get evidence in favor of one of the three regions
     from simulated data for 10\gap000 randomly chosen sets of probabilities 
     in each of the three regions.}  
\centerline{%
  \begin{tabular}{lcc@{\quad}c@{\quad}c}\hline\hline
   mock-true && \multicolumn{3}{c}{number of cases with} \\[-0.6ex]
   probabilities && \multicolumn{3}{c}{evidence in favor of region} \\[-0.6ex]
   in region    &\quad\quad& QM only & both & LHV only  \\
\hline
  QM only   && 8\gap809 & 1\gap278 &      0   \\
  both      &&      0   & 8\gap365 & 1\gap635 \\
  LHV only  &&      0   &     145  & 9\gap855 \\
\hline\hline
  \end{tabular}
}
\end{table}

For the $10\gap000$ mock-true $p$s in the ``QM only'' region we get evidence
in favor of this region in $8\gap809$ cases and $1\gap278$ times in favor of 
the ``both'' region, but never for the ``LHV only'' region; there are
$(8\gap809+1\gap278)-10\gap000=87$ instances with evidence in favor of the
``QM only'' region and also the ``both'' region. 
The mock-true $p$s in the ``both'' and the ``LHV only'' regions never result in
evidence in favor of the ``QM only'' region.
It follows that there is no bias in the prior toward the ``QM only''
region and, therefore, our finding that the actual data give strong evidence
in favor of this region, and against the other two, cannot be attributed to a
bias in the prior. 

One can read Table~\ref{tbl:BoulderData-nobias} as stating the (approximate)
probabilities of finding evidence for one region, conditioned on the true $p$
being from the same or another region.
It is tempting to invoke Bayes's theorem and convert this into the probability
that the true $p$ is in a certain region, given that we have evidence in favor
of one of the regions.
We resist this temptation because a correct application of Bayesian inference
requires correctly assigned prior probabilities, which we consciously choose not
to have \cite{correct-Bayes}.

\begin{table}[b]
\caption{\label{tbl:BoulderData-1or3or7pulses}%
Boulder data: Recorded counts in the runs one trigger signal per trial (top)
for ${N=175\gap647\gap100}$ trigger signals (top);
with three trigger signals per trial, for ${N=527\gap164\gap272}$ trigger
signals (middle); and with seven trigger signals per trial, for
${N=1\gap244\gap205\gap032}$ trigger signals (bottom)~\cite{Boulder-data}.}
\centerline{\begin{tabular}{c@{\quad}c@{\quad}c@{\quad}c@{\quad}c}
\hline\hline\rule{0pt}{12pt}%
$S$ & $n_{++}^{(S)}$ & $n_{+0}^{(S)}$ & $n_{0+}^{(S)}$ & $n_{00}^{(S)}$\\
\hline 
$ab$                 & 1\gap257 &      629 &      600 & 43\gap917\gap556\\
$ab^{\prime}$         & 1\gap417 &      554 & 4\gap549 & 43\gap908\gap718\\
$a^{\prime}b$         & 1\gap281 & 4\gap341 &      554 & 43\gap899\gap021\\
$a^{\prime}b^{\prime}$ &       11 & 5\gap640 & 6\gap030 & 43\gap894\gap942
\\[-1ex]
&\multicolumn{4}{c}{\hrulefill}\\
$ab$               & 3\gap800 &  1\gap936 &  1\gap812 & 131\gap804\gap979\\
$ab^{\prime}$        & 4\gap091 &  1\gap682 & 13\gap781 & 131\gap777\gap583\\
$a^{\prime}b$        & 3\gap853 & 12\gap840 &  1\gap669 & 131\gap749\gap135\\
$a^{\prime}b^{\prime}$ &       60 & 16\gap614 & 17\gap934 & 131\gap752\gap503
\\[-1ex]
&\multicolumn{4}{c}{\hrulefill}\\
$ab$               & 8\gap820 &  4\gap640 &  4\gap433 & 311\gap074\gap665\\
$ab^{\prime}$        & 9\gap512 &  3\gap963 & 32\gap709 & 310\gap997\gap997\\
$a^{\prime}b$        & 9\gap237 & 30\gap040 &  4\gap037 & 310\gap933\gap331\\
$a^{\prime}b^{\prime}$ &      159 & 38\gap632 & 42\gap034 & 311\gap010\gap823\\
\hline\hline
\end{tabular}}
\end{table}

\subsection{Trials with one, three, or seven trigger signals}%
\label{sec:Boulder-1or3or7}%
The event counts in the runs of the Boulder experiment with one, three, or seven
trigger signals per trial are given in
Table~\ref{tbl:BoulderData-1or3or7pulses}.
In total they have ${N=175\gap647\gap100}$,
${N=527\gap164\gap272}$, and ${N=1\gap244\gap205\gap032}$
events, respectively.
Table~\ref{tbl:BoulderData-SC1or3or7} reports the prior and posterior contents
of the three regions, and the data of the corresponding bias checks are given in
Table~\ref{tbl:BoulderData-nobias1or3or7}. 

\newcommand{\ThisSpace}{\rule{6pt}{0pt}}
\begin{table}
\caption{\label{tbl:BoulderData-SC1or3or7}%
  Boulder data: Prior and posterior contents
  of the three regions for the experimental data of Table
  \ref{tbl:BoulderData-1or3or7pulses}; for $\gamma=0.000\gap722$. } 
\centerline{\begin{tabular}{cccccccccc}
\hline\hline\rule{0pt}{0pt}%
 &\ThisSpace& \multicolumn{2}{c}{one trigger} 
 &\ThisSpace& \multicolumn{2}{c}{three triggers} 
 &\ThisSpace& \multicolumn{2}{c}{seven triggers}\\[-2pt]
 &\ThisSpace& \multicolumn{2}{c}{per trial} 
 &\ThisSpace& \multicolumn{2}{c}{per trial} 
 &\ThisSpace& \multicolumn{2}{c}{per trial}\\
region && prior & post. && prior & post. && prior & post.\\ \hline
 QM only && 0.000\gap6  & 1 && 0.000\gap6  & 1 && 0.000\gap6   & 1\\
  both   && 0.502\gap5  & 0 && 0.502\gap5  & 0 && 0.502\gap5   & 0\\
LHV only && 0.497\gap0  & 0 && 0.496\gap9  & 0 && 0.496\gap9   & 0\\ 
\hline\hline
\end{tabular}}
\end{table}

\begin{table}[b]
  \caption{\label{tbl:BoulderData-nobias1or3or7}%
   Boulder data: 
   The analog of Table~\ref{tbl:BoulderData-nobias} for the data of
   Table~\ref{tbl:BoulderData-1or3or7pulses}. 
   Bias checks for the runs with one (top),  three (middle), and seven (bottom)
   trigger signals per trial; for $\gamma=0.000\gap722$.}
\centerline{%
  \begin{tabular}{lcc@{\quad}c@{\quad}c}\hline\hline
  mock-true && \multicolumn{3}{c}{number of cases with} \\[-0.6ex]
  probabilities  && \multicolumn{3}{c}{evidence in favor of region}\\[-0.6ex]
  in region    &\quad\quad& QM only & both & LHV only  \\
\hline
  QM only   && 9\gap106 & 1\gap239 &      0   \\
  both      &&      0   & 8\gap478 & 1\gap522 \\
  LHV only  &&      0   &    85    & 9\gap915 \\[-1ex] 
  &&\multicolumn{3}{c}{\hrulefill}\\
  QM only   && 8\gap898 & 1\gap229 &      0   \\
  both      &&      2   & 8\gap384 & 1\gap614 \\
  LHV only  &&      0   &   129    & 9\gap871 \\[-1ex] 
  &&\multicolumn{3}{c}{\hrulefill}\\
  QM only   && 8\gap808 & 1\gap249 &      0   \\
  both      &&      0   & 8\gap344 & 1\gap656 \\
  LHV only  &&      0   &    83    & 9\gap917 \\
\hline\hline
  \end{tabular}
}
\end{table}

The analysis of the data from these three runs confirms the conclusion of
Sec.~\ref{sec:Boulder-5}: 
While there is no bias in the prior toward the ``QM only'' region, all of the
posterior is in this region; therefore, each run by itself provides strong
evidence in favor of the ``QM only'' region and against the other two.
Recall, however, what we noted at the beginning of Sec.~\ref{sec:Boulder},
namely that the four different runs of the Boulder experiment do not use
independently recorded raw data.

\section{The other three experiments}\label{sec:3others}
\subsection{The Vienna experiment}\label{sec:Vienna}
With reference to the parameters in Table~\ref{tbl:4exp}, we recall that the
Vienna experiment is similar to the Boulder experiment, with a larger number
of trigger signals and a larger nominal value of $\gamma$ while the other
parameters have about the same values.
The datasets numbered 6, 7, and 8 are available for 
evaluation~\cite{Vienna-data}; see Table~\ref{tbl:ViennaData}.

\begin{figure}
  \centerline{\includegraphics{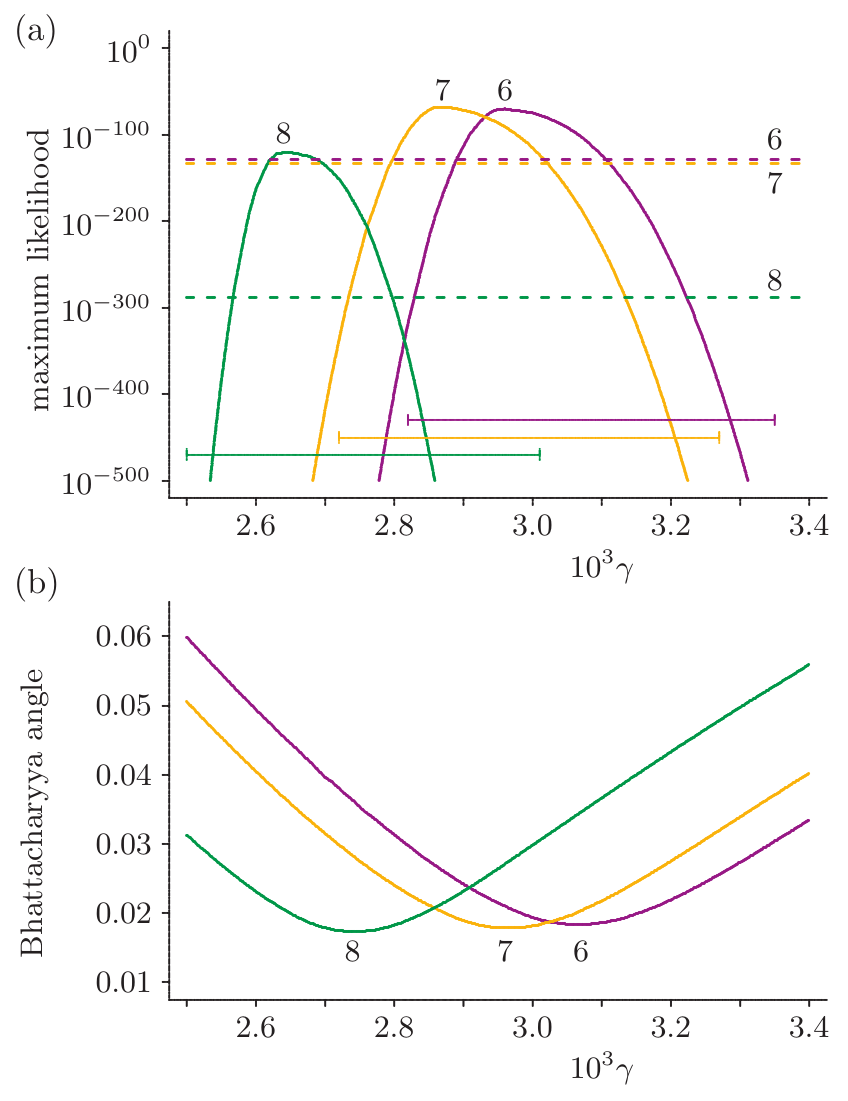}}  
  \caption{\label{fig:MaxLik2}%
    Vienna data: 
    (a) For ${0.002\gap5\leq\gamma\leq0.003\gap4}$, the solid curves
    graph the maximum values of $L(D|p)$ over the QM-per\-mitted
    probabilities (inside the symbolic ellipse in \Fig{regions}) for dataset~6
    (purple), dataset~7 (orange), and dataset~8 (green); the dashed
    lines mark the corresponding maximum values of $L(D|p)$ over the
    LHV-permitted probabilities (inside the symbolic triangle in
    \Fig{regions}).
    The values for $\gamma=0.003\gap5$ are far outside the range of the plot:
    $3.9\times10^{-1286}$ for dataset~6,
    $5.5\times10^{-1875}$ for dataset~7, and
    $2.8\times10^{-13060}$ for dataset~8; see also
    Table~\ref{tbl:ViennaData-MaxLik}.
    The horizontal bars show the respective ranges of $\gamma$ values for
    which the probabilities of the QM-MLE violate a Bell inequality of the
    Eberhard kind; these ranges correspond to the grey strip in \Fig{MaxLik1}.
    \newline(b) For the same $\gamma$ range, the curves graph the Bhattacharyya
    angle between the $p_{\alpha\beta}^{(S)}$s of the target state and those
    of the QM-MLE.
    The angle is smallest for ${\gamma=0.003\gap07}$, $0.002\gap96$, and
    $0.002\,74$ for dataset~6, dataset~7, and dataset~8, respectively.}
\end{figure}

Mindful of the lesson learned about the crucial importance of the value of
$\gamma$, we estimate its value from the data by the QSE procedure described
in Sec.~\ref{sec:gamma}.
The result of maximizing $L(D|p)$ over the QM-permissible or the
LHV-permissible sets of probabilities are reported in \Fig{MaxLik2}(a) for the
relevant range of $\gamma$ values, the analog of \Fig{MaxLik1} for the Boulder
data. 
We observe that there is not one common best-guess value for $\gamma$, as is
the case for the Boulder experiment, but we have three different optimal
$\gamma$ values for the three Vienna datasets, namely 
$\gamma_6=0.002\gap96$ for dataset~6, $\gamma_7=0.002\gap87$ for dataset~7, and
$\gamma_8=0.002\gap64$ for dataset~8.
The value of $\gamma=0.003\gap5$ in Table~\ref{tbl:4exp} is not an option;
it is also outside the ranges of $\gamma$ values where the probabilities of the
QM-MLE violate a Bell inequality of the Eberhard kind. 
The maximum values of the likelihood compiled in
Table~\ref{tbl:ViennaData-MaxLik} demonstrate the case: 
For these best-guess values of $\gamma$ the observed data are much more likely
for QM than for LHV, by many orders of magnitude.
If we were to take $\gamma=0.003\gap5$ seriously, the data would be much much
more likely for LHV than for QM, by more than 1\gap100 orders of magnitude for
dataset~6, and more than 12\gap700 for dataset~8.

\begin{table}[b]
\caption{\label{tbl:ViennaData}
Vienna data: Recorded counts in dataset 6 (top), dataset 7 (middle), and
dataset 8 (bottom), with 
${N=3\gap843\gap698\gap536}$, ${N=3\gap502\gap784\gap150}$, 
and ${N=9\gap994\gap696\gap192}$ 
trigger signals, respectively~\cite{Vienna-data}.}
\centerline{\begin{tabular}{c@{\quad}c@{\quad}c@{\quad}c@{\quad}c}
\hline\hline\rule{0pt}{12pt}%
$S$ & $n_{++}^{(S)}$ & $n_{+0}^{(S)}$ & $n_{0+}^{(S)}$ & $n_{00}^{(S)}$\\
\hline 
$ab$               & 159\gap976 &  83\gap743 &  86\gap270 & 960\gap597\gap110\\
$ab^{\prime}$        & 166\gap265 &  78\gap407 & 370\gap252 & 960\gap099\gap455\\
$a^{\prime}b$        & 179\gap813 & 482\gap787 &  66\gap435 & 960\gap381\gap485\\
$a^{\prime}b^{\prime}$ &   9\gap354 & 655\gap290 & 525\gap368 & 959\gap756\gap526
\\[-1ex]
&\multicolumn{4}{c}{\hrulefill}\\
$ab$               & 141\gap439 &  73\gap391 &  76\gap224 & 875\gap392\gap736\\
$ab^{\prime}$        & 146\gap831 &  67\gap941 & 326\gap768 & 874\gap976\gap534\\
$a^{\prime}b$        & 158\gap338 & 425\gap067 &  58\gap742 & 875\gap239\gap860\\
$a^{\prime}b^{\prime}$ &   8\gap392 & 576\gap445 & 463\gap985 & 874\gap651\gap457
\\[-1ex]
&\multicolumn{4}{c}{\hrulefill}\\
$ab$               & 377\gap000 &      192\gap092 & 202\gap207 & %
                     2\gap497\gap825\gap793\\
$ab^{\prime}$        & 387\gap481 &      182\gap789 & 858\gap681 & %
                     2\gap496\gap663\gap605\\
$a^{\prime}b$        & 422\gap674 & 1\gap119\gap219 &     156\gap022 & %
                     2\gap497\gap626\gap620\\
$a^{\prime}b^{\prime}$ &  22\gap502 & 1\gap519\gap578 & 1\gap223\gap007 & %
                     2\gap495\gap916\gap922\\
\hline\hline
\end{tabular}}
\end{table}

\makeatletter\global\advance\@colroom10pt\relax\set@vsize\makeatother

What is said in Sec.~\ref{sec:gamma} is equally fitting here:
(i) The computation of p-values for the purpose of refuting the LHV hypothesis
does not require an accurate value of~$\gamma$, but our Bayesian analysis
needs an accurate value for the QM probabilities.
(ii) When considered from the QM perspective, the actually observed data are
not typical at all for $\gamma=0.003\gap5$.

In \Fig{MaxLik2}(b) we show the Bhattacharyya angle $\phi_{\mathrm{B}}^{\ }$
between the probabilities of the target state and those of the QM-MLE.
Analogous remarks to those about \Fig{MaxLik1} apply here as well: 
The angle is small for the $\gamma$ values for which the corresponding QM
likelihood is large, and this supports our conclusion that
$\gamma_6=0.002\gap96$, $\gamma_7=0.002\gap87$, and
$\gamma_8=0.002\gap64$ are values with much better justification than
$\gamma=0.003\gap5$.

\begin{table}
  \caption{\label{tbl:ViennaData-MaxLik}%
   Vienna data: Maximum values of the likelihood $L(D|p)$ 
   for QM-permissible probabilities with
   selected values of $\gamma$, and for LHV-permissible probabilities.}
  \centerline{\begin{tabular}{lcclll}\hline\hline%
   & $\gamma$ &\quad& dataset~6 & dataset~7 & dataset~8\\ \hline
   & $0.002\gap64$ && $7.9\times10^{-1503}$  
                               & $5.3\times10^{-732}$ 
                               & $5.1\times10^{-121}$    \\
   & $0.002\gap87$ && $2.4\times10^{-170}$  
                               & $8.5\times10^{-69}$ 
                               & $1.8\times10^{-569}$    \\ 
   & $0.002\gap96$ && $6.3\times10^{-71}$  
                               & $1.7\times10^{-91}$ 
                               & $3.4\times10^{-1097}$    \\ 
\begin{turn}{90}\makebox[0pt][l]{%
   \underline{\rule{15pt}{0pt}QM\rule{15pt}{0pt}}}\end{turn}
   & $0.003\gap5\phantom{0}$  && $3.9\times10^{-1286}$  
                               & $5.5\times10^{-1875}$ 
                               & $2.8\times10^{-13060}$    \\[-1ex]
         &&&\multicolumn{3}{c}{\hrulefill}\\
   \multicolumn{2}{l}{LHV, any $\gamma$}&   & $4.7\times10^{-129}$  
                             & $5.5\times10^{-134}$ 
                             & $3.4\times10^{-289}$    \\
  \hline\hline
  \end{tabular}}
\end{table}

While slightly different values for $\gamma_6$, $\gamma_7$, and $\gamma_8$ are
equally acceptable, we emphasize that there is no common value.
It appears that the intensity of the pump laser, whose pulses trigger the
generation of down-converted photon pairs with entangled polarization qubits,
was largest for dataset~6 and smallest for dataset~8, with correspondingly
different $\gamma$ values \cite{Wengerowsky-private}.
We proceed under the assumption that the value of $\gamma$ was stable enough
during each run that it is reasonable to apply a single effective
$\gamma$ value to each of the three datasets~\cite{drift}.

\begin{table}
\caption{\label{tbl:ViennaData-SC}%
  Vienna data: Prior and posterior contents
  of the three regions for the experimental data of Table
  \ref{tbl:ViennaData} with the respective best-guess values for $\gamma$.
  As in Table~\ref{tbl:BoulderData-SC} above and also in 
  Table~\ref{tbl:DelftMunichData-SC} below, the prior
  contents have sampling errors of no consequence; see
  Sec.~\ref{sec:SamplingError}.} 
\centerline{\begin{tabular}{cclcclcclc}
\hline\hline\rule{0pt}{0pt}%
 && \multicolumn{2}{c}{data~set~6} 
 && \multicolumn{2}{c}{data~set~7} 
 && \multicolumn{2}{c}{data~set~8}\\[-0.8ex]
 && \multicolumn{2}{c}{${\gamma=0.002\gap96}$} 
 && \multicolumn{2}{c}{${\gamma=0.002\gap87}$} 
 && \multicolumn{2}{c}{${\gamma=0.002\gap64}$}\\[-0.8ex]
region &\quad\quad& prior & post. &\qquad& %
prior & post.&\qquad& prior & post.\\ \hline
 QM only && 0.001\gap8  & 1 && 0.001\gap9   & 1 && 0.001\gap8 & 1\\
  both   && 0.501\gap8  & 0 && 0.501\gap7   & 0 && 0.501\gap8 & 0\\
LHV only && 0.496\gap4  & 0 && 0.496\gap4   & 0 && 0.496\gap4 & 0\\ 
\hline\hline
\end{tabular}}
\end{table}

When checking the datasets of the Vienna experiment for evidence in
favor of or against the regions symbolized in \Fig{regions}, we find the prior
and posterior contents of Table~\ref{tbl:ViennaData-SC}.
Once more we have strong evidence in favor of the ``QM only'' region and
against the other two.
This conclusion is what the data tell us.
It is not a consequence of a biased prior, as is demonstrated by the
due-diligence bias check documented in Table~\ref{tbl:ViennaData-nobias} which
uses ten thousand mock-true sets of probabilities for each region and each
data set.

\begin{table}
  \caption{\label{tbl:ViennaData-nobias}%
   Vienna data: 
   The analogs of Tables~\ref{tbl:BoulderData-nobias} and
   \ref{tbl:BoulderData-nobias1or3or7}, here for the data of
   Table~\ref{tbl:ViennaData}. 
   Bias checks for dataset~6 (top), dataset~7 (middle), and dataset~8
   (bottom), for the respective best-guess values of $\gamma$.}
\centerline{%
  \begin{tabular}{lcc@{\quad}c@{\quad}c}\hline\hline
  mock-true && \multicolumn{3}{c}{number of cases with} \\[-0.6ex]
  probabilities  && \multicolumn{3}{c}{evidence in favor of region}\\[-0.6ex]
  in region    &\quad\quad& QM only & both & LHV only  \\
\hline
  QM only   &&  9\gap225 &   778     &  0  \\
  both      &&     0     & 8\gap545  &  1\gap455  \\
  LHV only  &&     0     &    94     &  9\gap906\\[-1ex] 
  &&\multicolumn{3}{c}{\hrulefill}\\
  QM only   &&  9\gap229 &   775    &   0  \\
  both      &&     1     & 8\gap341 & 1\gap658 \\
  LHV only  &&     0     &    137    & 9\gap863 \\[-1ex]
  &&\multicolumn{3}{c}{\hrulefill}\\
  QM only   &&  9\gap264  &    736    &  0  \\
  both      &&     1      & 8\gap599  & 1\gap400  \\
  LHV only  &&     0      &     82    & 9\gap918 \\
\hline\hline
  \end{tabular}
}
\end{table}

\makeatletter\global\advance\@colroom12pt\relax\set@vsize\makeatother

\subsection{The Delft  and Munich experiments}\label{sec:Delft+Munich}
As a consequence of exploiting other physical systems, the parameters in
Table~\ref{tbl:4exp} are quite different for the Delft and Munich experiments
than for the Vienna and Boulder experiments.
In particular, there is ${\gamma=1}$ in the Delft and Munich experiments, and
no estimation of $\gamma$ is called for, and also 
${\eta^{\ }_{\mathrm{A}},\eta^{\ }_{\mathrm{B}}\lesssim1}$ are more advantageous
values, whereas there are much fewer trigger signals, which is a drawback.

\begin{table}
\caption{\label{tbl:DelftMunichData}
Delft and Munich data: 
Recorded counts in run~1 (top left) and run~2 (top right) of the Delft
experiment, with  ${N=245}$ and ${N=228}$ trigger signals, 
respectively~\cite{Delft-data};
and in run~1 (middle) and run~2 (bottom) of the Munich experiment, with  
${N=27\gap885}$ and ${N=27\gap683}$ 
trigger signals, respectively~\cite{Munich-data}.
The marked difference in the relative frequencies for ${S=a'b}$ and 
${S=a'b'}$ in the two runs of the Munich experiment originates in the 
use of two different target states.} 
\centerline{\begin{tabular}{lc@{\quad}c@{\quad}c@{\quad}c@{\quad}c}
\hline\hline\rule{0pt}{12pt}%
& $S$ & $n_{++}^{(S)}$ & $n_{+0}^{(S)}$ & $n_{0+}^{(S)}$ & $n_{00}^{(S)}$\\
\hline 
&$ab$               & 23$|$21 & 3$|$7 & 4$|$3 & 23$|$21\\  
&$ab^{\prime}$        & 33$|$25 & 11$|$2\phantom{0} & 5$|$4 & 30$|$23\\ 
&$a^{\prime}b$        & 22$|$19 & 10$|$11 & 6$|$6 & 24$|$23\\  
\begin{turn}{90}\makebox[0pt][l]{%
   \underline{\rule{12.5pt}{0pt}Delft\rule{12.5pt}{0pt}}}\end{turn}
&$a^{\prime}b^{\prime}$ & 4$|$5 & 20$|$24 & 21$|$23 & \phantom{0}6$|$11
\\[-1ex]
&&\multicolumn{4}{c}{\hrulefill}\\ %
&$ab$               &   778    & 2\gap621 & 2\gap770 &   804\\
&$ab^{\prime}$        &   809    & 2\gap629 & 2\gap708 &   816\\
&$a^{\prime}b$        &   873    & 2\gap686 & 2\gap644 &   730\\
&$a^{\prime}b^{\prime}$ & 2\gap696 &   966    &   902    & 2\gap453  
\\[-1ex]
&&\multicolumn{4}{c}{\hrulefill}\\
&$ab$               &    817   & 2\gap596 & 2\gap873 &   742\\ 
&$ab^{\prime}$        &    696   & 2\gap570 & 2\gap788 &   772\\
&$a^{\prime}b$        & 2\gap783 &   787    &    840   & 2\gap503\\
\begin{turn}{90}\makebox[0pt][l]{%
   \underline{\rule{38.5pt}{0pt}Munich\rule{38.5pt}{0pt}}}\end{turn}
&$a^{\prime}b^{\prime}$ &    865   & 2\gap620 & 2\gap640 &   791\\  
\hline\hline
\end{tabular}}
\end{table}

The data from two runs each for the Delft and the Munich experiments are
available for evaluation~\cite{Delft-data,Munich-data}; 
see Table~\ref{tbl:DelftMunichData}. 
We begin the evaluation with a maximization of the likelihood $L(D|p)$ over
the set of QM-permissible probabilities and also over the set of
LHV-permissible probabilities. 
As the entries in Table~\ref{tbl:DelftMunichData-MLE} show, the observed data
are much more likely for QM-permissible than for LHV-permissible
probabilities, with ratios between $7.2$ and $6.5\times10^{15}$.
Since the two runs of the Delft experiment had exceptionally few events,
less than one-hundredth of the counts in the Munich experiment,
we also combined the data of the two Delft runs into one larger set with
${N=473}$ trigger signals (``run 1\&2''), and that is included in
Table~\ref{tbl:DelftMunichData-MLE} as well \cite{Delft-3rd}. 
Not only is there no corresponding need to combine the data of the two runs
of the Munich experiment, this would not be proper to begin with, 
as there were two different target states; see Table~\ref{tbl:DelftMunichData}. 

\begin{table}
\caption{\label{tbl:DelftMunichData-MLE}%
   Delft and Munich data:
   Maximum values of the likelihood $L(D|p)$ 
   for QM-permissible probabilities and for LHV-permissible probabilities.
   The last column gives the ratios of the two maximum-likelihood values.}
  \centerline{%
  \begin{tabular}{l@{\quad}c@{\quad}c@{\quad}c}\hline\hline
    &QM&LHV&ratio\\ \hline
  Delft run~1    & $2.95\times10^{-16}$ & $4.10\times10^{-17}$ &  $ 7.2$ \\ 
  Delft run~2    & $8.20\times10^{-15}$ & $2.63\times10^{-15}$ &  $ 3.1$ \\
  Delft run~1\&2 & $1.22\times10^{-17}$ & $1.04\times10^{-18}$ &  $ 12$ \\[0.5ex]
  Munich run~1   & $2.17\times10^{-30}$ & $5.27\times10^{-34}$ %
                 & $4.1\times10^{3}$ \\
  Munich run~2   & $2.88\times10^{-31}$ & $4.46\times10^{-47}$ %
                 & $6.5\times10^{15}$ \\
  \hline\hline
  \end{tabular}}
\end{table}

\begin{table*}
\caption{\label{tbl:DelftMunichData-SC}%
  Delft and Munich data: 
  Prior and posterior contents of the three regions for the experimental 
  data of Table~\ref{tbl:DelftMunichData}.}
\centerline{\begin{tabular}{ccllcllcllclcclc}
\hline\hline\rule{0pt}{0pt}%
 && \multicolumn{2}{c}{Delft, run~1} 
 && \multicolumn{2}{c}{Delft, run~2} 
 && \multicolumn{2}{c}{Delft, run 1\&2}
 && \multicolumn{2}{c}{Munich, run~1} 
 && \multicolumn{2}{c}{Munich, run~2} \\[-0.7ex]
region &\quad\quad& prior & posterior &\qquad& %
prior & posterior &\qquad& prior & posterior &\qquad\qquad& %
prior & posterior &\qquad& prior & posterior\\ \hline
 QM only && 0.151\gap2  & 1.000\gap000 && %
            0.151\gap2  & 0.999\gap980 && %
            0.151\gap3  & 1.000\gap000 && %
            0.076\gap9  & 1 && %
            0.077\gap0  & 1\\
  both   && 0.362\gap7  & 1.2\ten{7} && %
            0.362\gap7  & 7.0\ten{6} && %
            0.362\gap8  & 6.7\ten{9}&& %
            0.426\gap7  & 0 && %
            0.426\gap6  & 0\\
LHV only && 0.486\gap0  & 4.5\ten{8} && %
            0.486\gap0  & 1.3\ten{5} && %
            0.486\gap0  & 1.6\ten{10}&& %
            0.496\gap4  & 0 && %
            0.496\gap4  & 0\\ 
\hline\hline
\end{tabular}}
\end{table*}

\begin{figure}[b]
  \centerline{\includegraphics{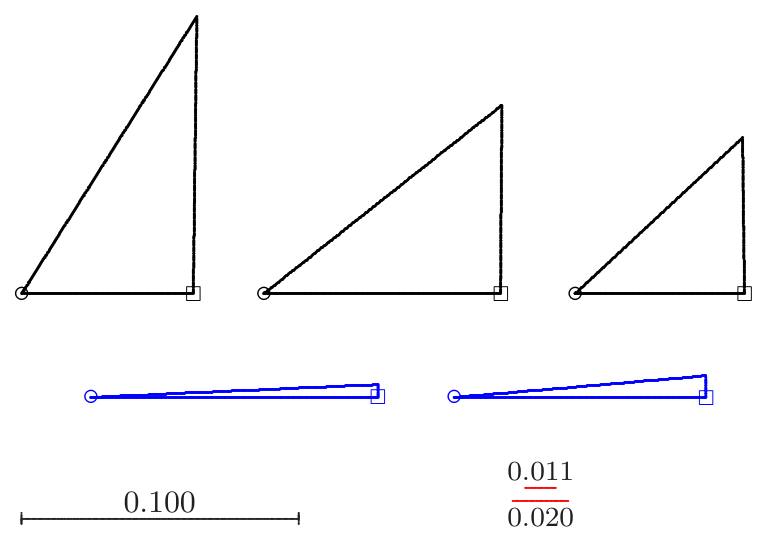}}  
  \caption{\label{fig:DelftMunichTriangles}%
  Delft and Munich data: Probability-space triangles.
  The corners of the triangles indicate the sets of probabilities for the
  respective target state (circle), QM-MLE (square), and the relative
  frequencies observed (unmarked corner).
  The black triangles at the top are for runs~1, 2, and 1\&2 of the Delft
  experiment (left to right) and the blue triangles below are
  for the two runs of the Munich experiment. 
  The lengths of the sides of the triangles are the Bhattacharyya angles
  between the probabilities at the corners, on the indicated scale.}
\end{figure}

The Bhattacharyya angle is a genuine distance between two sets of probabilities.
Therefore, the three angles between the relative frequencies, 
the $p^{(S)}_{\alpha\beta}$s of the target states, and those of the QM-MLE,
constitute the sides of a triangle. 
In \Fig{DelftMunichTriangles} we show the three triangles for runs~1, 2, and
1\&2 of the Delft experiment at the top (in black) and the two
triangles for the runs of the Munich experiment below (in blue). 
These are five independent drawings with no relations among them, except that
the scale is the same, defined by the reference line for an angle of $0.100$.
In each triangle, the circle marks the target state, the square marks the
QM-MLE, and the third corner is for the relative frequencies.

As it should be, the QM-MLE is always nearer to the relative frequencies than
the target state, somewhat nearer for the Delft runs, much nearer for the
Munich runs. 
The Bhattacharyya angles between the probabilities of the target state and
those of the QM-MLE for the data of the Boulder and Vienna experiments,
evaluated for the respective best-guess value for $\gamma$,
are markedly smaller, namely $0.011$ for the Boulder experiment, and
$0.020$ for the Vienna experiment. 
The red-line stretches indicate these distances in \Fig{DelftMunichTriangles}; 
they are to be compared with the distances between circles and squares.

This comparison indicates that the precision with which the intended target
state is realized in the Boulder and Vienna experiments is noticeably better
than that in the Delft and Munich experiments, which certainly did also
succeed with good precision. 
While this observation has no bearing on our conclusions, it does illustrate
what the data tell us when putting questions to them.   

For the data in Table~\ref{tbl:DelftMunichData} we find the prior and
posterior contents of the three regions reported in
Table~\ref{tbl:DelftMunichData-SC}, including also the combined Delft data
(``run 1\&2'').
When comparing the prior contents in Table~\ref{tbl:DelftMunichData-SC} with
those in Tables~\ref{tbl:BoulderData-SC}, \ref{tbl:BoulderData-SC1or3or7}, and
\ref{tbl:ViennaData-SC}, we notice that the ``QM only'' region has a much
larger prior content for the Delft and Munich data than for the Boulder and
Vienna data. 

This difference results predominantly from an increase of the ``QM only''
fraction and a decrease of the ``both'' fraction in the QM sample (see
Sec.~\ref{sec:QM-p}).
This different distribution in turn originates mainly in the larger detection
efficiencies $\eta^{\ }_{\mathrm{A}}$ and $\eta^{\ }_{\mathrm{B}}$ and, to a
lesser extent, also in the larger angles $\theta^{\ }_{\mathrm{A}}$ and 
$\theta^{\ }_{\mathrm{B}}$; see Table~\ref{tbl:4exp}.
These parameters directly enter the QM probabilities of \Eq[Eqs.~]{B5} and
\Eq[]{B6} but not the LHV probabilities of \Eq{B10}.
Notice also that the ``QM only'' region for the Delft experiment has
about twice the prior content of that for the Munich experiment,
mostly a consequence of the imperfections mentioned in
note~\cite{Munich-dark}. 
By contrast, the very substantial difference in the value of $\gamma$ is of
little consequence because a change in $\gamma$ leads to an overall scaling of
the regions without changing their relative size, as is most clearly seen when
we look at the accessible region of just one of the four 3-simplices (recall
that ${p^{(S)}_{00}\geq1-\gamma}$). 

The posterior contents for the Munich data are as extreme as we observed 
for the Boulder and the Vienna data: 
Only the ``QM only'' region has posterior content.
Just like the data of the Boulder and Vienna experiments do, the data of the
Munich experiment give very strong evidence in favor of the ``QM only'' region
and against the other two regions. 

While the data of the Delft experiment also give evidence of the same kind,
the evidence is less strong than in the other experiments, simply because
there are so many fewer events.
Nevertheless, the evidence against LHV is strong, certainly stronger than the
p-value of 0.039 suggests~\cite{Delft}.
This p-value is below the conventional 0.05 threshold which is taken to
imply evidence against; see also~\cite{Benjamin+71:17}.
We note, however, that even if a p-value was above any such threshold, then
this could not be taken as evidence for the hypothesis in question and that
is in sharp contrast to the approach we follow here.

\begin{table}
  \caption{\label{tbl:DelftMunichData-nobias}%
   Delft and Munich data: 
   The analog of Tables~\ref{tbl:BoulderData-nobias}, 
   \ref{tbl:BoulderData-nobias1or3or7}, and \ref{tbl:ViennaData-nobias},
   here for the bias check on the priors for the data of
   Table~\ref{tbl:DelftMunichData}.
   One thousand mock-true $p$s are used for each region and run.}
\centerline{%
  \begin{tabular}{lllcc@{\quad}c@{\quad}c}\hline\hline
  &&mock-true && \multicolumn{3}{c}{number of cases with} \\[-0.6ex]
  &&probabilities  && \multicolumn{3}{c}{evidence in favor of region}\\[-0.6ex]
  &&in region    &\quad\quad& QM only & both & LHV only  \\
\hline
  &&QM only   && 971 & 153 &   0 \\
  &&both      &&  65 & 810 & 216 \\
  &&LHV only  &&   8 &  48 & 958 \\[-1ex] 
  &&&&\multicolumn{3}{c}{\hrulefill}\\
  &&QM only   &&  973 & 176 &   4 \\
  &&both      &&   67 & 801 & 236 \\
  &&LHV only  &&    5 &  46 & 961 \\[-1ex] 
  &&&&\multicolumn{3}{c}{\hrulefill}\\
  &&QM only   &&  979 & 124 &   0 \\
  &&both      &&   47 & 852 & 177 \\
  \begin{turn}{90}\makebox[0pt][l]{%
   \underline{\rule{53.5pt}{0pt}Delft\rule{53.5pt}{0pt}}}\end{turn}
  &\begin{turn}{90}\makebox[0pt][l]{%
   \underline{\rule{0pt}{0pt}run~1\&2\rule{0pt}{0pt}}\rule{14pt}{0pt}%
   \underline{\rule{5.5pt}{0pt}run~2\rule{5.5pt}{0pt}}\rule{14pt}{0pt}%
   \underline{\rule{5.5pt}{0pt}run~1\rule{5.5pt}{0pt}}}\end{turn}
  &LHV only  && 1 & 25 & 980\\[-1ex] 
  &&&&\multicolumn{3}{c}{\hrulefill}\\
  &&QM only   &&  982 &  25 &   0 \\
  &&both      &&    5 & 860 & 136 \\
  &&LHV only  &&    0 &  12 & 988 \\[-1ex] 
  &&&&\multicolumn{3}{c}{\hrulefill}\\
  &&QM only   &&  988 &  22 &   0 \\
  &&both      &&    3 & 865 & 133 \\
  \begin{turn}{90}\makebox[0pt][l]{%
   \underline{\rule{24.5pt}{0pt}Munich\rule{24.5pt}{0pt}}}\end{turn}
 &  \begin{turn}{90}\makebox[0pt][l]{%
   \underline{\rule{5.5pt}{0pt}run~2\rule{5.5pt}{0pt}}\rule{14pt}{0pt}%
   \underline{\rule{5.5pt}{0pt}run~1\rule{5.5pt}{0pt}}}\end{turn}
&LHV only  &&   0 & 9 & 991\\
\hline\hline
  \end{tabular}
}
\end{table}

Here, too, the analysis would be incomplete without a confirmation that there
is no bias in the prior.
Table~\ref{tbl:DelftMunichData-nobias} shows how often we find evidence in
favor of the three regions when simulating data for one-thousand mock-true $p$s
from each of the regions.
As discussed in Sec.~\ref{sec:bias-check}, the obvious difference between this
Table and Tables~\ref{tbl:BoulderData-nobias},
\ref{tbl:BoulderData-nobias1or3or7}, and \ref{tbl:ViennaData-nobias} originates
in the smaller counts of event (Munich) or the much smaller ones (Delft).
Even so, although it can happen more easily here that we find evidence in
favor of the ``QM only'' region from data for probabilities from another
region, there is no procedural bias for the ``QM only'' region.

\section{Discussion and conclusion}\label{sec:discuss}
Our analysis is based entirely on the event counts in 
Tables~\ref{tbl:BoulderData-5pulses}, \ref{tbl:BoulderData-1or3or7pulses}, 
\ref{tbl:ViennaData}, and \ref{tbl:DelftMunichData}, with no other information
about the recorded data. 
Therefore, our analysis must assume that for each run of one of the four
experiments, the corresponding list of counts is a sufficient statistic.
In particular, this brings up the issues of notes \cite{runs-test} and
\cite{drift}, namely that the sequences of detected events do not exhibit
correlations that should not be there, so that the experiments can be reliably
evaluated with the parameters in Table~\ref{tbl:4exp}, even if we found it
necessary to estimate the trigger--to--qubit-pair conversion probabilities
from the data themselves (Boulder and Vienna).  

Therefore, we do not worry about the so-called ``memory loophole'' --- the
notion that the LHV could keep track of past outcomes and adjust the
probabilities for future ones in the most deceiving way. 
Nor do we entertain scenarios in which the detectors communicate with
each other through unknown means (dark-matter waves perhaps?) with, again,
fitting adjustments for future outcomes.
Rather, we take for granted that the data have been thoroughly checked for
correlations that would result from such mechanisms, and that none were found.

As explained in the Introduction, our choice of prior does not follow the
rules of proper Bayesian reasoning.
Our prior, defined by the sampling algorithm described in
Sec.~\ref{sec:prior}, ignores the rules deliberately.
It does not take into account any prior knowledge we have about the
experimental situation --- that there is strong prior evidence for QM and none
at all for LHV, and that the experimenters have the skills to build the
apparatus as specified.  
Instead, our prior leans heavily toward LHV --- the prior content of the ``LHV
only'' region is larger in all samples, sometimes much larger, than that of
the ``QM only'' region --- and that there really is no procedural bias for QM
is demonstrated by our bias checks, one such test of the prior used in each
experimental context. 
The bias checks are a priori and do not depend on the observed data.

Even with all this support extended to LHV, the data provide very strong
evidence in favor of QM and against LHV.
This is especially convincing in the face of the procedural advantage given to
LHV. 

In closing, we wish to remind the reader that nonquantum formalisms with LHV
do not amount to a serious alternative to the QM description.
Successful LHV accounts of any recorded data have always been limited to very
particular experimental situations and relied on a case-by-case ad-hoc
reasoning. 
The plethora of phenomena that are correctly accounted for by QM are simply
beyond the reach of LHV.
Yet, even within this limited context in which LHV have a slim chance of
success they have been refuted for good.

\section{Outlook}
While the ``QM vs LHV case'' is at the center stage of this work, there is a
more general lesson here about the use of Bayesian methods when asking what
evidence is provided by empirical data.
On the conceptual side, there is the Bayesian principle of evidence:
We have evidence in favor of an alternative if it is more probable after the
data are available than before; and there is evidence against the alternative
if the data render it less probable.
On the procedural side, there is the systematic checking of the chosen prior
for an unwanted bias in order to ensure that the conclusions are not
predetermined.

This approach is applicable to many situations.
Let us mention just one.
Suppose you have a source of qutrit pairs and you want to verify that it
prepares the pairs in a state with bound entanglement~\cite{Sentis+4:18}.
You could then collect tomographic data and determine whether there is
evidence in favor of the set of bound-entangled states, or against it.

\acknowledgments
We sincerely thank Ronald Hanson (Delft), Anton Zeilinger and S\"oren 
Wengerowsky (Vienna), Krister Shalm (Boulder), and Harald Weinfurter,
Kai Redeker, and Wenjamin Rosenfeld (Munich) for most valuable correspondence.   
Enlightening exchanges with Jacek Gruca, Marek \.Zukowski, and Dagomir
Kaszlikowski about LHV are gratefully acknowledged.
We owe many thanks to David Nott for numerous helpful discussions about
Bayesian reasoning, to Hui Khoon Ng for asking probing questions, and to both
of them for their much appreciated advice.
This work is funded by the Singapore Ministry of Education (partly through the
Academic Research Fund Tier 3 MOE2012-T3-1-009) and the National Research
Foundation of Singapore.


\begin{thebibliography}{99}
\bibitem{strength}
See Chapter 4 in Ref.~\cite{Evans:15}.
We note that it is customary to measure the strength of evidence by the size
of the so-called ``Bayes factor'' but this practice is questionable in view of
the Jeffreys--Lindley paradox \cite{Jeffreys:39,Lindley:57} and other
deficiencies.
\bibitem{fn:Munich-eta}
In the detection stage of the Munich experiment, atoms are ionized, and the
ions and electrons are detected separately, with efficiencies of
$0.90\cdots0.94$ for the ion detection and $0.75\cdots0.90$ for the electron
detection.
Since detection is successful if either an ion or an electron or both are
found, this results in an overall detection efficiency between 0.975 and
0.994.
The lower limit is displayed in Table~\ref{tbl:4exp} and used for the data
evaluation reported in Sec.~\ref{sec:Delft+Munich};
other values could be used for $\eta_{\mathrm{A}}^{\ }$ and
$\eta_{\mathrm{B}}^{\ }$ equally well, with no change in our final conclusions.
\bibitem{runs-test}
We trust that runs tests have confirmed this statistical independence.
We cannot perform runs tests ourselves on all data sets
as we do not know the sequence of detection events for some of them.
For all experiments evaluated, we only explore the total counts of recorded
detection events (Tables~\ref{tbl:BoulderData-5pulses},
\ref{tbl:BoulderData-1or3or7pulses}, \ref{tbl:ViennaData},
and \ref{tbl:DelftMunichData}), not the order in which they were recorded.
The list of counts is a sufficient statistic if the events are statistically
independent, and only then.
\bibitem{Signaling}
The no-signaling conditions of \Eq[Eqs.~]{B2} and \Eq[]{B3} matter.
If we maximize $L(D|p)$ by only imposing the four constraints of \Eq{B1}, we
obtain $2.12\times10^{-46}$, more than seven times the value with the
no-signaling conditions enforced.
\bibitem{correct-Bayes}
By a correct application of Bayesian inference we mean the use of a prior
that fully reflects our strong beliefs concerning the truth of QM and we have
purposefully not done that here but rather chose to be much more even-handed.
\bibitem{drift}
A remark similar to that in note~\cite{runs-test} applies:
We cannot test this assumption on the basis of the data in
Table~\ref{tbl:ViennaData} as one would need to compare the data recorded at
different intervals of the data-taking period.
We trust that the data have been tested for drifts in the experimental
parameters and none were found within each of the three datasets.
\bibitem{Delft-3rd}
There is also a ``third run'' of the Delft experiment with a different target
state as run~1 and run~2 and a total of $N=72$ trigger signals
\cite{Delft-2nd}. 
We do not include these data in our analysis.
\end{thebibliography}
\end{document}